\documentclass[runningheads]{llncs}

\usepackage{booktabs}
\usepackage[utf8]{inputenc}
\usepackage{booktabs,caption}
\usepackage[flushleft]{threeparttable}
\usepackage{color}

\usepackage{amssymb,amsfonts,amsmath,amsthm,amscd,dsfont}
\usepackage[margin=1in]{geometry}
\usepackage{graphicx,float,epsfig}
\usepackage{wrapfig}
\usepackage{hyperref}
\usepackage{algorithm,algorithmic}
\usepackage{bbm}
\usepackage{authblk}
\usepackage{csquotes}
\usepackage{enumitem}

\usepackage{color}
\usepackage{xcolor}

\DeclareMathAlphabet{\mathpzc}{OT1}{pzc}{m}{it}

\newtheorem{propo}{Proposition}[]
\newtheorem{thm}[propo]{Theorem}

\def\prob{{\mathbb P}}
\def\tx{\tilde{x}}
\def\tX{\tilde{X}}
\def\reals{\mathbb{R}}

\def\E{\mathbb{E}}

\def\Rc{r_c}
\def\Rgi{r_g}
\def\Rm{r_M}
\def\ones{\mathds{1}}

\renewcommand\footnotemark{}

\title{Compounding of  Wealth in Proof-of-Stake Cryptocurrencies}

\author{
Giulia Fanti$^\ddagger$, 
Leonid Kogan$^\star$, 
Sewoong Oh$^\dagger$, Kathleen Ruan$^\ddagger$, Pramod Viswanath$^\dagger$, Gerui Wang$^\dagger$
\thanks{Email: gfanti@andrew.cmu.edu, lkogan2@mit.edu, swoh@illinois.edu, kruan@andrew.cmu.edu, pramodv@illinois.edu, geruiw2@illinois.edu. 
}
}
\institute{
$^\ddagger$Carnegie Mellon University, $^\star$Massachusetts  Institute of Technology, $^\dagger$University of Illinois  Urbana-Champaign
}

\begin{document}

\maketitle

\begin{abstract}
Proof-of-stake (PoS) is a promising approach for designing efficient blockchains, where block proposers are randomly chosen with probability proportional to their stake.
A primary concern with PoS systems is the ``rich getting richer" phenomenon, whereby wealthier nodes are more likely to get elected, and hence  reap the block reward, making them even wealthier. 
In this paper, we introduce the notion of equitability, which quantifies how much a proposer can amplify her stake compared to her  initial investment. 
Even with everyone following protocol (i.e., honest behavior), we show that existing methods of allocating block rewards lead to poor equitability, as does initializing systems with small stake pools and/or large rewards relative to the stake pool.
We identify a \emph{geometric} reward function, which we prove is maximally equitable over all choices of reward functions under honest behavior and bound the deviation for strategic actions; 
the proofs involve  the study  of optimization problems 
and stochastic dominances of 
  P\'{o}lya urn processes, 
and are of independent mathematical interest. These results allow us to provide a systematic framework to choose the parameters of a  practical incentive system for  PoS cryptocurrencies. 
\end{abstract}

\section{Introduction}
A central problem in blockchain systems is that of block proposal: how to choose which \emph{block}, or set of transactions, should be appended to the global blockchain next. 
Many blockchains use a proposal mechanism by which one node is randomly selected as leader (or \emph{block proposer}). 
This leader gets to propose the next block in exchange for a token reward---typically a combination of transaction fees and a freshly-minted \emph{block reward}, 
which is chosen by the system designers.
These reward mechanisms incentivize nodes to participate in the block proposal procedure, and are therefore critical to the security and liveness of the system. 

Early cryptocurrencies, including Bitcoin, overwhelmingly used a leader election mechanism called \emph{proof of work} (PoW).
Under PoW, all nodes execute a computational puzzle.
The node who solves the puzzle first is elected leader; she proves her leadership by broadcasting a solution to the puzzle before the other nodes.
Over the years, PoW showed itself to be extremely robust to security threats, 
but also extremely energy-inefficient. 
The Bitcoin network alone is estimated to use more energy than some developed nations \cite{energy}. 

An appealing alternative to PoW is called \emph{proof-of-stake} (PoS). 
In PoS, proposers are not chosen according to their computational power, but according to the stake they hold in the cryptocurrency. 
For example, if Alice has 30\% of the tokens, she is selected as the next proposer with probability 0.3.
Although the idea of PoS is both natural and energy-efficient, the research community is still grappling with how to design a PoS system that provides security while also incentivizing nodes to act as network validators. 
Part of incentivizing validators is simply providing enough reward (in expectation) to compensate their resource usage. 
However, it is also important to ensure that validators are treated fairly compared to their peers. 
In other words, they cannot only be compensated adequately on average; the variance also matters.

This observation is complicated in PoS systems by a key issue that does not arise in PoW systems: \emph{compounding}.
Compounding means that whenever a node (Alice) earns a proposal reward, that reward is added to her account, which increases her chances of being elected leader in the future, and increases her chances of reaping even more rewards. 
This leads to a rich-get-richer effect, causing dramatic concentration of wealth. 

\begin{figure}[h]
	\begin{centering}
	\includegraphics[width=2.5in]{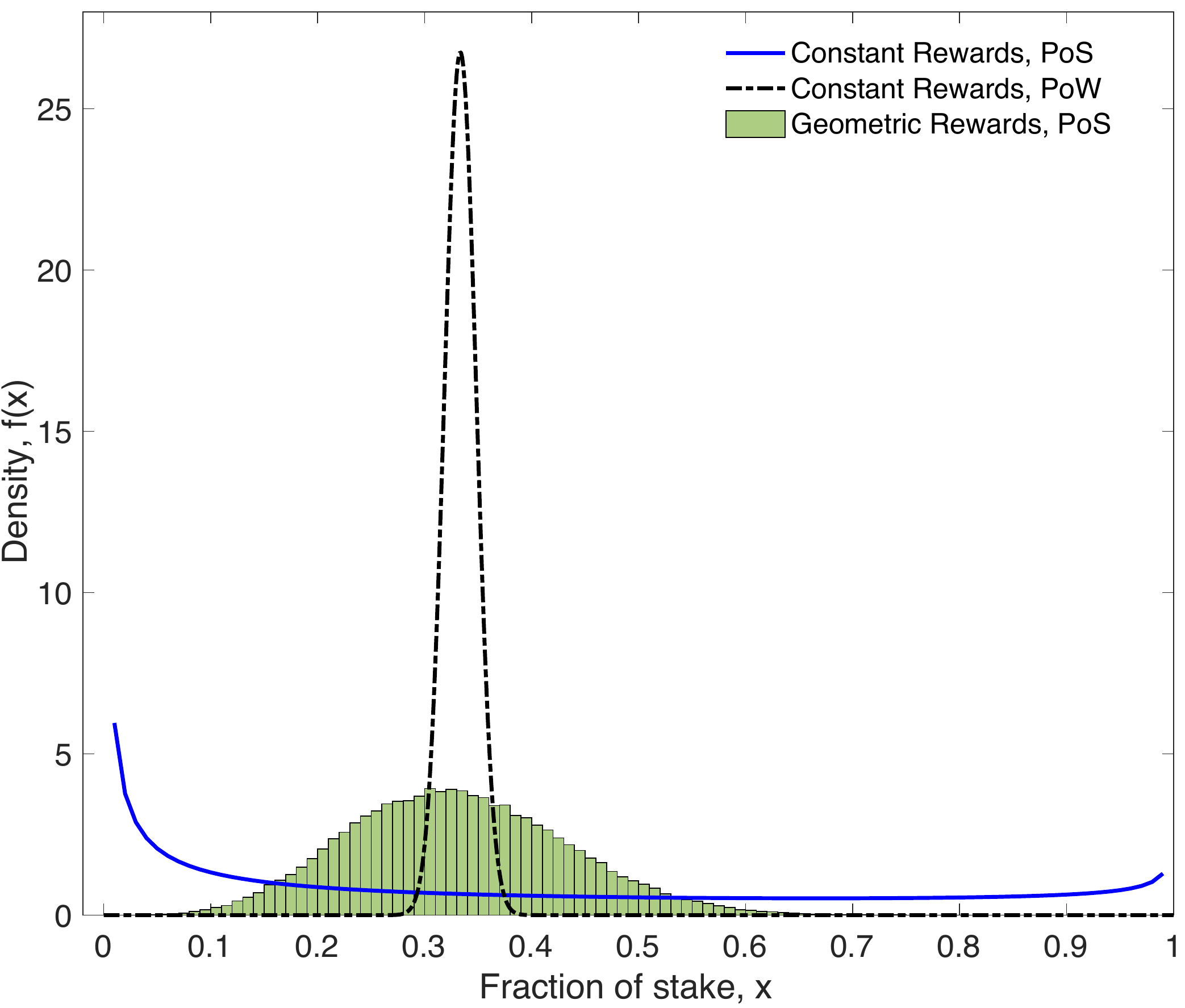}
	\caption{
		Fractional stake distribution of a party that starts with ${1}/{3}$ of 
		the stake in a system initialized with Bitcoin's financial parameters. 
		Results of geometric reward PoS and constant reward PoW are shown after $T=1,000$ blocks. 
		Typical constant reward PoS systems suffer from the compounding effect, 
		which can be significantly mitigated by using the proposed geometric reward function. 
	}
	\label{fig:beta}
	\end{centering}
\end{figure}

For example, consider what would happen if Bitcoin were a PoS system.
Bitcoin started with
an initial stake pool of $50$ BTC, and the block reward was fixed at $50$ BTC/block for several years.
Under these conditions, suppose a party $A$ starts with $\frac{1}{3}$ of the stake. 
Using a basic PoS model described in Section \ref{sec:model}, $A$'s stake would evolve according to a standard P\'olya urn process \cite{johnson1977urn},  converging almost surely to a random variable with distribution Beta$(\frac{1}{3},\frac{2}{3})$ \cite{mahmoud2008polya}, (blue solid line in Figure \ref{fig:beta}). 
In this example,  compounding gives $A$ a high probability of accumulating a stake fraction near 0 or 1. 
This is highly undesirable because the proposal incentive mechanism should not unduly amplify or shrink one party's fraction of stake.
Notice that this is \emph{not} caused by an adversarial or strategic behavior, but simply due to the 
 {\em randomness} in the PoS protocol, {\em combined with compounding}.

In PoW, on the other hand, the analogue would be for party $A$ to hold $1/3$ of the computational power. 
In that case, $A$'s stake after $T$ blocks would be instead binomially distributed with mean $({50\times T})/{3}$ (black dashed line in Figure \ref{fig:beta}).
Notice that the binomial (PoW) stake distribution concentrates around ${1}/{3}$ as $T\to \infty$, so if $A$ contributes ${1}/{3}$ of stake at the beginning, she also reaps ${1}/{3}$ of the rewards in the long term. 
A natural question is whether we can achieve this PoW baseline distribution in a PoS system with compounding.

We study this question from the perspective of the block reward function. 
Most cryptocurrencies today use a \emph{constant block reward} function like Bitcoin's, which remains fixed over a long timespan (e.g., years).
We ask how a PoS system's choice of block reward function can affect concentration of wealth, 
and whether one can achieve the PoW baseline stake distribution simply by changing the block reward function.
This paper has five main contributions:
\begin{enumerate}
\item We define the \emph{equitability} of a block reward function, which intuitively
captures how much 
the fraction of total stake belonging to 
a node 
can grow or shrink (under that block reward function), compared to the node's initial investment.  
An equitable block reward scheme should limit this variability.
This metric allows us to quantitatively compare reward functions.
\item We introduce an alternative block reward function called the \emph{geometric reward function}, 
whose rewards increase geometrically over time.
We show that it is the most equitable PoS block reward function, 
by showing that it is the unique solution to an 
optimization problem on the second moment of a time-varying urn process;
this optimization may be of independent interest in the applied probability community.
We further show that geometric rewards exhibit a number of desirable properties, including stability of rewards in fiat value over time. 
We note that despite optimizing equitability, geometric rewards do not achieve the PoW baseline stake distribution -- this is the {\em inherent} price paid by the efficiency afforded by PoS compared to (the energy inefficient) PoW. 
The green histogram in Figure \ref{fig:beta} illustrates the empirical, simulated stake distribution when geometric rewards are awarded over a duration of $1,000$ blocks, and the total rewards are the same as in the PoW example (i.e., $50\times 1,000$ units). 
These simulations are run over 100,000 trials.

\item Borrowing ideas from resource pooling in PoW systems, 
a plausible strategy of participants  with small stakes in a PoS system is 
to collectively form larger stake pools. 
We quantify exactly the gain of such stake pool formation in terms of equitability, 
which proves that 
participating in a stake pool  can significantly 
reduce the compounding effect of a PoS system.

\item 
We study the effects of strategic behavior (e.g. selfish mining) on the rich-get-richer phenomenon. 
We find that in general, compounding can exacerbate the efficacy of strategic behavior compared to PoW systems. 
However, these effects can be partially mitigated by carefully choosing the amount of block reward dispensed over some time period relative to the initial stake pool size.  

\item Our analyses of the equitability of various reward functions 
provide guidelines for choosing system parameters---including the  
initial token pool size and the total rewards to dispense in  a given time interval---to ensure equitability under a given block reward function. In particular, we show that cryptocurrencies that start with large initial stake pools (relative to the block rewards being disseminated) can mitigate the concentration of wealth, both for constant and geometric reward schemes.

\end{enumerate}

The rest of this paper is organized as follows. 
In Section \ref{sec:model}, we present our model, and discuss the relation between it and real PoS cryptocurrencies. 
We also precisely define the constant and geometric block reward functions. 
In Section \ref{sec:honest}, we compare honest and geometric block reward schemes, showing that geometric rewards exhibit optimal equitability over \emph{all} reward schemes. The resulting design decisions in choosing practical parameters of PoS system block reward schemes are discussed in Section~\ref{sec:equitability_guideline}. 
We use Section \ref{sec:strategic} to study the effects of strategic behavior on equitability --   we find that neither constant nor geometric rewards provide robustness against selfish-mining-type attacks. The desired robustness to strategic behavior in PoS systems is perhaps designed via suitable incentive (and disincentive) mechanisms, as discussed in Section \ref{sec:system}. 

%
%
%
%
%
%
%
%
%


%

\subsection{Related work}\label{sec:related}
The potential of poor equitability of PoS systems has been explored in some detail in 
recent forum and blog posts in the cryptocurrency community  \cite{pos_rich,pos_rich2,pos_rich3},
but no research has formally or quantitatively studied it (to the best of our knowledge). 
In this work, we quantify concentration of wealth through a new metric called equitability, which enables us to mathematically  compare PoS to PoW, as well as different block reward schemes. 
As we discuss in Section \ref{sec:model}, equitability is closely tied to the variance of a block reward scheme.
Thus far, researchers and practitioners have reduced variance in block rewards through two main approaches: 
 pooling resources (e.g., mining or stake pools) and
 proposing new protocols for disseminating block rewards. 

Resource pooling is a common phenomenon in cryptocurrencies. 
For example, since PoW mining requires substantial computational resources, few nodes are independently able to mine profitably. 
Mining pools democratize this process by allowing many nodes to participate in mining, while also sharing block rewards among those nodes \cite{schrijvers2016incentive,eyal2018majority}. 
In PoS systems, the analogous concept is stake pooling, where nodes aggregate their stake under a single node;  block rewards are shared across the pool. 
Like mining pools, stake pools allow less 
wealthy players to participate in network maintenance. 
In Section \ref{sec:pool}, we show how much one can gain by participating in a 
stake pool in terms of equitability, and show that 
the proposed geometric reward function is still the most equitable even if some of the parties involved are forming stake pools. 

Consider a different scenario where 
some reward is deterministically dispensed to 
every participant of a PoS system at each block proposal according to some predeclared rules.  
In particular, the block proposer is treated no differently from any other participating party. 
There is no randomness in this system and hence no compounding effect. 
Under this assumptions, \cite{brunjes} studies  a problem of an organic stake formation, 
where any participant is allowed to create a stake pool, where 
she acts as a leader of the pool at some cost. 
The PoS system designer can choose 
a reward function $r(a,b):\reals\times\reals\to \reals\times \reals$ to be appleid to each pool, 
where $a$ is the total stake of the pool, $b$ is the stake that the leader holds, and $r_1(a,b)$ 
is to be shared equally  over all participants of the pool according to their stake, 
and $r_2(a,b)$ is awarded to the leader. 
The goal of the system designer is to organically form 
a fixed target number $k$ of stake  pools, by choosing the reward function.
Our work differs from \cite{brunjes} in three main respects:  
(1) While our work aims to optimize equitability, \cite{brunjes} aims to incentivize the formation of a target number of mining pools. 
(2) We study the effects of compounding on concentration of wealth, whereas \cite{brunjes} does not model compounding. 
(3) We study the dynamic setting  as opposed to static setting. 


A second class of approaches for reducing variance actually changes the protocol for block reward allocation. 
Our work falls into this category.
Two main examples of this approach are Fruitchains \cite{fruitchains}, which spread block rewards evenly across a sequence of block proposers, and Ouroboros \cite{kiayias2017ouroboros}, which rewards nodes for being part of a block formation committee, even if they do not contribute to block proposal.
Both of these approaches were proposed in order to provide incentive-compatibility for block proposers; they do not explicitly aim to reduce the variance of rewards. 
However, they implicitly reduce variance by spreading rewards across multiple nodes, thereby preventing the randomized accumulation of wealth. 
In our work, instead of changing how block rewards are disseminated, we change the block reward function itself. 


\section{Models and Notation} 
\label{sec:model}

We provide a probabilistic model for the evolution of the stakes under a PoS system, and 
introduce a measure of fairness, we call {\em equitability}.

\subsection{A Simple PoS model} 
We begin with a model of a  chain-based proof-of-stake system 
with $m$ parties: $\mathcal A = \{A_1, \ldots, A_m\}$.
We assume that all parties keep all of their stake in the \emph{proposal stake pool}, which is a pool of tokens that is used to choose the next proposer. 
We consider a discrete-time system, $n=1,2,\ldots, T$, where each time slot corresponds to the addition of one block to the blockchain.
In reality, new blocks may not arrive at perfectly-synchronized time intervals, but we index the system by block arrivals. 
For any integer $x$, we use the notation $[x] := \{1,2,\ldots, x\}$.
For all $i\in [m]$, let $S_{A_i}(n)$ denote the total stake held by party $A_i$ in the proposal stake pool at time $n$.
We let $S(n) = \sum_{i=1}^m S_{A_i}(n)$ denote the total stake in the proposer stake pool at time $n$, 
and $v_{A_i}(n)$ denotes the \emph{fractional stake} of node $A_i$  at time $n$: 
$$
v_{A_i}(n) = \frac{S_{A_i}(n)}{S(n)}.
$$
For simplicity, we normalize the initial stake pool size to $S(0)=1$; 
this is without loss of generality as the  random process is homogeneous 
in scaling both the rewards and the initial stake by a constant. 
Each party starts with $S_{A_i}(0)=v_{A_i}(0)$ fraction of the original stake.

At each time slot $n \in [T]$, the system chooses a proposer node $W(n) \in \mathcal A$ such that 
\begin{eqnarray}
 W(n)=
\begin{cases}
A_1  & w.p.\quad v_{A_1}(n) \\
\ldots \\
A_m  & w.p. \quad v_{A_m}(n).
\end{cases}
\label{eq:Polya}
\end{eqnarray}
Upon being selected as a proposer, $W(n)$ appends a \emph{block}, or set of transactions, to the \emph{blockchain}, which is a sequential list of blocks held by all nodes in the system. 
As compensation for this service, $W(n)$ receives a \emph{block reward} of $r(n)$ stake, which is immediately added to its allocation in the proposer pool.
That is, 
$$S_{W(n)}(n+1) = S_{W(n)}(n) + r(n).$$
The reward $r(n)$ is freshly-minted at each time step, so it causes the total number of tokens to grow.
We assume the total reward dispensed in time period $T$ is fixed, such that $\sum_{n=1}^T r(n)=R$.

\subsection{Modeling Assumptions}
This model implicitly makes several assumptions. 
The first is that we assign a single leader (proposer) per time slot. 
Many cryptocurrencies have leader election protocols that allow more than one proposer to be chosen per time slot (e.g., Bitcoin, PoSv3, Snow White). 
If two leaders are elected at time $n$, for example, then each leader can append its block to one block at height $n-1$; here the \emph{height} of a block is its index in the blockchain. 
However, in these systems, only one leader can win the block reward since only one fork of the blockchain ultimately gets adopted.
Assuming the final winner is chosen uniformly at random from the set of selected leaders, the dynamics of our Markov process remain unperturbed.

Other cryptocurrencies (e.g., Qtum, Particl) choose the next proposer(s) as a function of the time slot \emph{and} the preceding block. 
Again, this can lead to multiple proposers per time slot.  
This does not affect our results in the honest setting (for the same reason as above), but it does impact strategic behavior.
In most blockchain systems, honest proposers always build on the head of the blockchain.
However, in systems where the proposer's identity depends on the previous block, a strategic node can increase its chance of being a leader by appending to a block that is not at the head of the blockchain.
If done repeatedly, the strategic player may eventually produce a chain that is longer than the honest chain (causing the honest nodes to switch over), which also contains mostly blocks belonging to the strategic player. 
This increases the player's reward, and is called a  \emph{grinding attack}.
Such PoS systems are more vulnerable to strategic behavior than the system we analyze, where proposer election is a function of only the time slot. 
Despite this, we find that our model is drastically vulnerable to strategic behavior.  
Hence, the problem can only be worse in blockchains that use block contents to choose the next leader. 
We discuss the implications of this in Section \ref{sec:system-strategic}.

We have also assumed that users always re-invest their rewards into the proposer stake pool. 
We maintain that this  is a reasonable assumption for two reasons: 
(1) In PoS systems where users explicitly deposit stake, existing implementations automatically deposit rewards back into the stake pool. 
For example, the reference implementation of Casper the Friendly Finality Gadget (a PoS finalization mechanism proposed for Ethereum) automatically re-allocates all rewards back into the deposited stake pool \cite{casper}.
(2) In other PoS systems, the stake pool is simply the set of all stake in the system, and is not separate from the pool of tokens used for transactions \cite{posv3}.
Hence as soon as a proposer earns a reward, that reward is used to calculate the next proposer (modulo some maturity  period); 
 the user is not actively re-investing block rewards---it just happens naturally.


Finally, we have chosen not to explicitly model node unavailability, e.g. due to hardware or network failures; in our context, node unavailability means that a selected proposer may forfeit its chance to propose, even though it was chosen.
Assuming such node failures occur i.i.d. across draws from the proposer pool, such events do not alter our model dynamics. 
If a proposer is offline, the selection process is simply re-run; 
the slot in question is given to the next node, which is again chosen proportionally to the stake allocation in the proposer pool.


\subsection{Block reward choices}
\begin{figure}[htbp]
\begin{minipage}{0.45\textwidth}
\begin{center}
\includegraphics[width=2.3in]{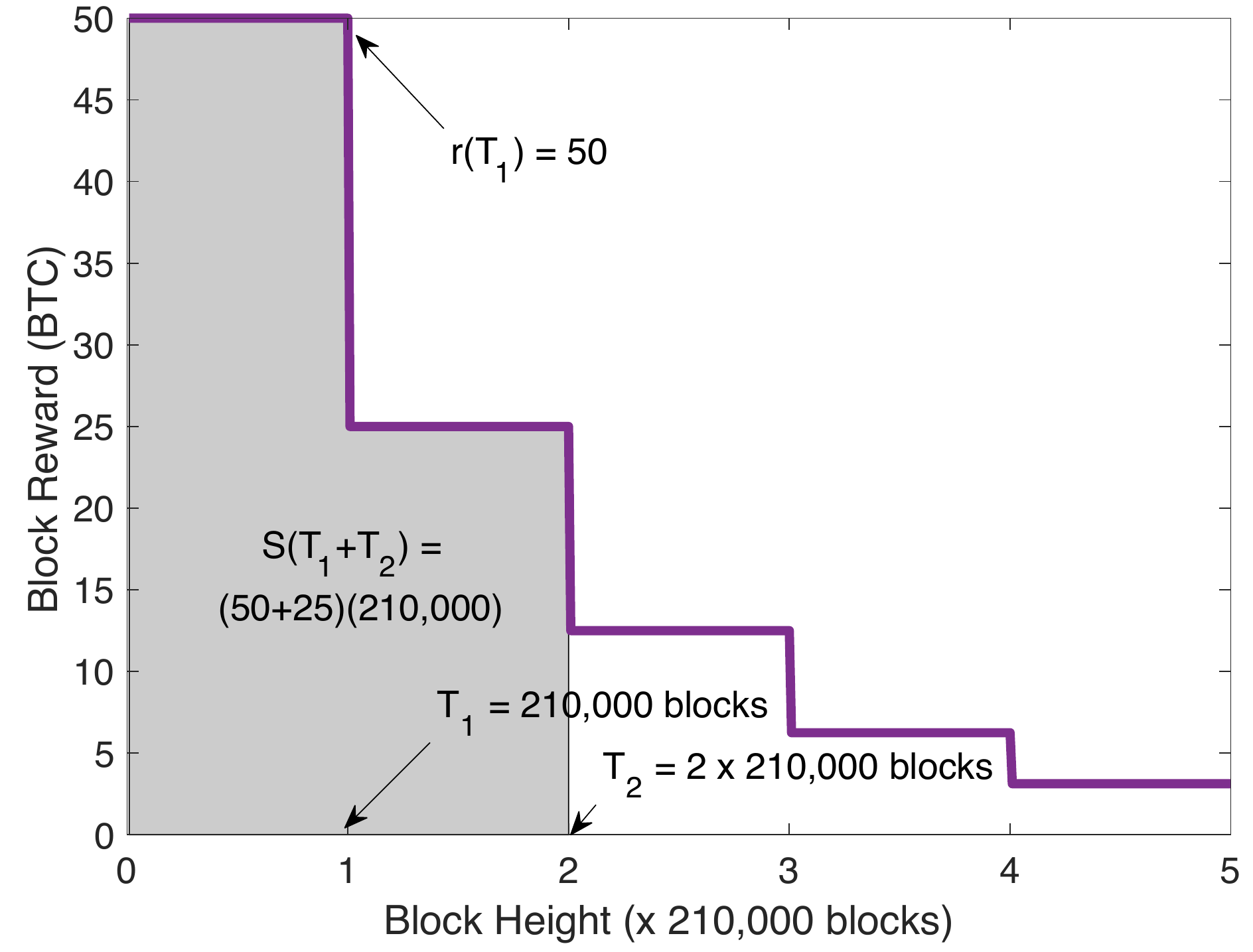}
\caption{Bitcoin block rewards as a function of block height. The area of the shaded region gives the total stake after $T_1+T_2$ time.}
\label{fig:btc-reward}
\end{center}
\end{minipage}
~~
\begin{minipage}{0.45\textwidth}
\begin{center}
\includegraphics[width=2.3in]{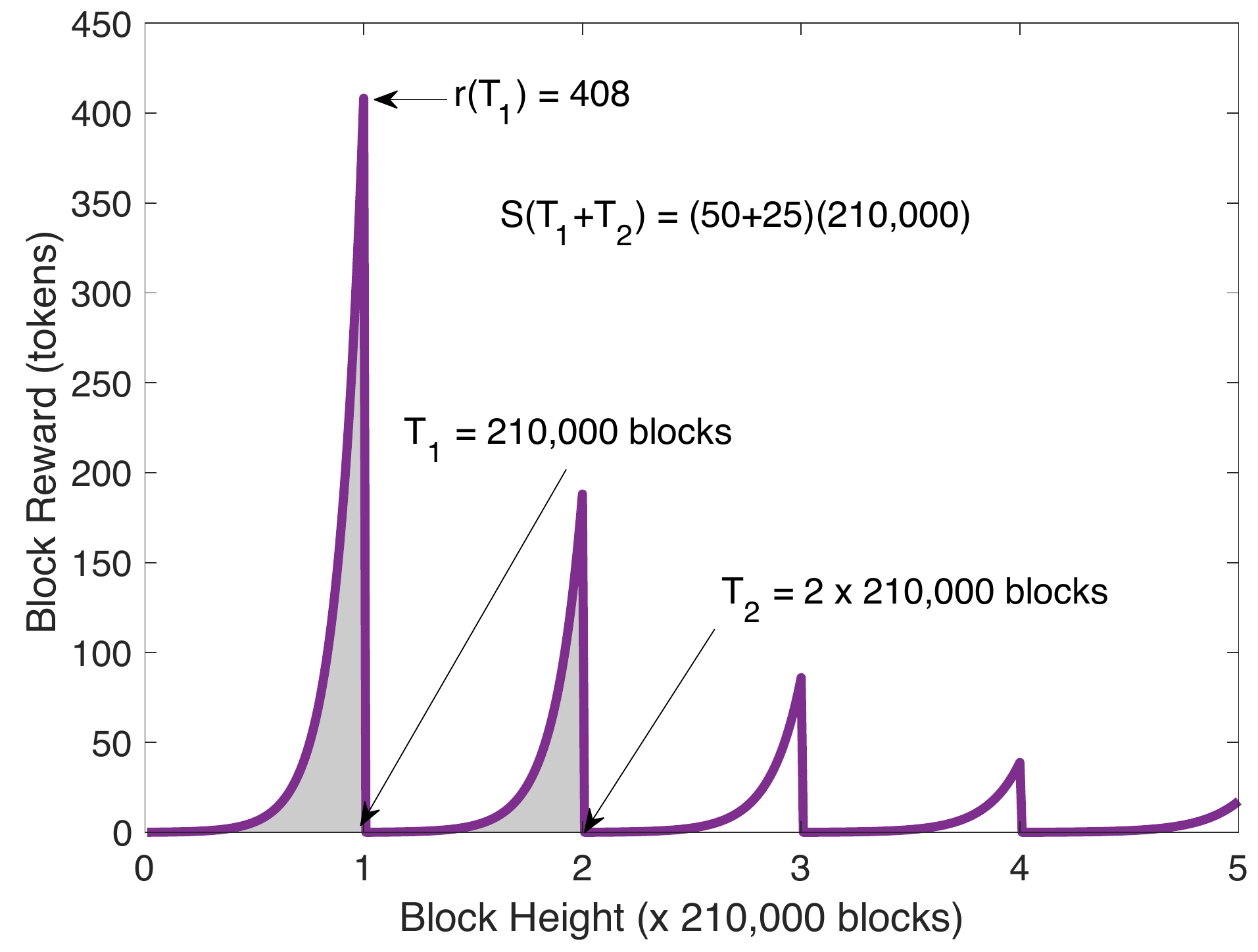}
\caption{Geometric block rewards as a function of block height, using Bitcoin-based $T_i$ and $R_i$ values from Figure \ref{fig:btc-reward}.}
\label{fig:geom-reward}
\end{center}
\end{minipage}
\end{figure}

Many cryptocurrencies have modeled their block reward strategy after Bitcoin's, which fixes the total supply of coins at about 21 million coins. 
To achieve this, block rewards are halved every 210,000 blocks (approximately four years) \cite{btc-schedule}.
In between halving events, the block reward remains constant. 
Figure \ref{fig:btc-reward} illustrates this reward schedule in terms of our notation;
if we let $T_i$ and $R_i$ denote the $i$th block interval and total reward, respectively, we can take $T_i=210,000$ blocks, and $R_i = 50\cdot \frac{1}{2^{i-1}} \cdot 210,000$. 
Several cryptocurrencies have similarly adopted block reward schemes that remain constant over extended periods of time, including Ethereum \cite{ethereumwiki}, ZCash \cite{hopwood2016zcash}, Dash \cite{duffield2015dash}, and Particl \cite{kaiserdecentralized}.
Note that choosing $T_i$ and $R_i$ is not our main focus; these parameters will likely be chosen based on economic considerations (we discuss  this in Sections \ref{sec:honest}, \ref{sec:equitability_guideline} and \ref{sec:strategic}).
Below we aim to provide guidelines on how to choose $r(n)$ once the $T_i$'s and $R_i$'s are fixed.

Other cryptocurrencies have experimented with the block reward function.
For example, Monero \cite{miller2017empirical} has a block reward that decays with each block; this is intended to be a continuous interpolation of Bitcoin's piecewise constant reward function \cite{van2013cryptonote}.
Peercoin \cite{anderson2016new}, one of the first PoS cryptocurrencies, chooses the next leader based on the age and quantity of stake associated with a given public key.
The PoS block reward is chosen as 1\% of the product of a public key's stake quantity and stake age \cite{king2012peercoin}; this differs from our model, where the block reward amount does not depend on the proposer who is selected.

\subsection{Systematic choice of reward functions}
In this paper, we revisit the question of how to choose $r(n)$. 
A key observation is that $r(n)$ is ultimately an incentive; it should compensate nodes for the resource cost of proposing blocks. 
Since this cost is roughly constant over time, many cryptocurrencies implicitly adopt the following maxim: 

\begin{displayquote}
\emph{On short timescales, each proposed block should yield the same block reward.}
\end{displayquote}
Notice that this maxim does not specify whether the value of a block reward is measured in {tokens} or in {fiat currency}. 
As illustrated earlier, most cryptocurrencies today measure value in tokens; that is, they give the same number of tokens for each block. 
We call this approach the \emph{constant block reward function}:
\begin{equation}
r_c(n) := \frac{R}{T}.
\label{eq:reward_const}
\end{equation}

A natural alternative is to measure the block reward's value in fiat currency. 
This approach depends closely on the cryptocurrency's valuation (and fluctuations thereof) over time interval $[T]$.
However, if we assume that the cryptocurrency's valuation is constant over $[T]$, then the resulting reward function should always give a constant fraction of the \emph{total} stake at each time slot.
We call this the \emph{geometric} reward function, defined  as follows:
\begin{equation}
r_g(n) := (1+R)^{\frac{n}{T}} - (1+R)^{\frac{n-1}{T}}.
\label{eq:reward_gi}
\end{equation}
Figure \ref{fig:geom-reward} shows geometric block rewards as a function of time if we use the same $T_i$'s and $R_i$'s as those in Figure \ref{fig:btc-reward}, which were tailored to Bitcoin's block reward schedule.
Note that a currency's valuation can change over the course of $T_i=210,000$ blocks; these parameters were chosen simply to ease the comparison with Figure \ref{fig:btc-reward}.
Our assumption that valuation remains constant can be enforced by choosing a small enough value of $T$. 
We discuss this parameter choice in Section \ref{sec:equitability_guideline}, but in short, we envision $T$ being on the order of a day.
Since $T$ is measured in units of blocks, this implies anywhere between thousands to tens of thousands of blocks per time interval.

\subsection{Equitability} 
\label{sec:equitability} 

To compare different reward functions, we define a metric called equitability.
Consider the stochastic dynamic of the fractional stake 
of a party $A$ 
that starts with $v_A(0)$ fraction of the initial total stake of $S(0)=1$. 
We denote the  fractional stake at time $n$ by $v_{A,r}(n)$, 
to make the dependence on the reward function explicit. 

One  desirable property of a PoS block reward function is that each node's 
fractional stake should remain constant over time 
in {\em expectation}.
If $A$ contributes 10\% of the proposal stake pool at the beginning of the time, 
then $A$ should reap 10\% of the total disseminated rewards on average. 
Since randomness in proposer elections is essential to current PoS systems, 
this cannot be ensured deterministically. 
Hence, a straw-man metric for quantifying fairness is the expected fractional stake at time $T$.
This metric turns out to be meaningless because most PoS systems elect a proposer (in Eq~\eqref{eq:Polya}) 
with probability proportional to the fractional stake; this approach ensures that each party's expected fractional reward is equal to its initial stake fraction, regardless of block reward function. 
Formally,  $\forall n\in[T]$, 
\begin{eqnarray}
	\label{eq:}
	\E[v_{A,r}(n) ] \;\; = \;\; v_A(0)\;. 
\end{eqnarray}
This follows from the law of total expectation and the fact that 
\begin{align*}
	&\E[ v_{A,r}(n)\,|\, v_{A,r}(n-1) = v] \\
	&\;\;\;\; = v  \frac{v\,S(n-1) + r(n-1)}{S(n)}+ \big(1-v \big) \frac{v\,S(n-1)}{S(n)} = v.
\end{align*}




Although all reward functions yield the same expected fractional stake, 
the choice of reward function can nonetheless dramatically change the distribution of the final stake, as seen in Figure \ref{fig:beta}. 
We therefore instead propose using the \emph{variance} of the final fractional stake, ${\rm Var}(v_{A,r}(T))$, as an equitability metric.  
Intuitively, smaller variance implies less uncertainty and therefore a higher level of equitability. 
We make this formal in the following definition. 

\begin{definition}
	\label{def:equitability}
	For a positive vector $\boldsymbol \varepsilon \in {\mathbb R}^m$, we say a reward function $r:[T]\to\reals^+$ 
	over $T$  time steps is {
	{\em $\boldsymbol \varepsilon$-equitable} for $\boldsymbol \varepsilon = [\varepsilon_1, \ldots, \varepsilon_m]$ where $\varepsilon_i>0$, if 
	\begin{eqnarray}
		\label{eq:equitability1}
		\frac{{\rm Var}(v_{A_i,r}(T))}{v_{A_i}(0)(1-v_{A_i}(0))} \;\; \leq \;\;  \varepsilon_i \;
	\end{eqnarray}
	for all $i\in [m]$.
	}

	For two reward functions 
	$r_1:[T] \to \reals^+$ and $r_2:[T] \to \reals^+$ 
	with the same total reward, $\sum_{n=1}^T r_1(n)=\sum_{n=1}^T r_2(n)$, 
	we say $r_1$ is more {\em equitable} than $r_2$ 
	{for player $i\in [m]$ if}
	\begin{eqnarray}
		\label{eq:equitability2}
		{\rm Var}\big(v_{A_i,r_1}(T)\big) \;\; \leq \;\;  {\rm Var}\big( v_{A_i,r_2}(T) \big) \;, 
	\end{eqnarray} 
	when both random processes start with the same initial fraction at each party of $v_{A_i}(0)$.
\end{definition}
The normalization in Eq.~\eqref{eq:equitability1} ensures the left-hand side is at most one, 
as we show in Remark \ref{rem:max_variance}.  
It also cancels out the dependence on the initial fraction $v_A(0)$
such that the left-hand side only depends on the reward function $r$ and the time $T$, 
 as shown in Lemma \ref{lem:lem_variance}. 

\begin{remark} When starting with an initial fractional stake $v_A(0)$, the maximum achievable variance is 
	\begin{eqnarray}
		\sup_{T\in {\mathbb Z}^+} \sup_{r} {\rm Var}(v_{A,r}(T)) \;\;=\;\; v_A(0)(1-v_A(0)) \;,
	\end{eqnarray}
	where the supremum is taken over all positive integer $T$ and reward function $r:[T]\to\reals^+$. 
	\label{rem:max_variance}
\end{remark}
\begin{proof}
	We first prove the converse, 
	${\rm Var}(v_{A,r}(T)) \leq v_A(0)(1-v_A(0))$ 
	for all $T$ and $r$. 
	This follows from the fact that 
	$\E[v_{A,r}(T)] = v_A(0)$, 
	and $v_{A,r}(T)$ is bounded below by zero and above by one.  
	Maximum variance is achieved when all probability mass 
	is concentrated on the boundary of zero and one. 
	
	We prove the achievability, by constructing a simple constant 
	reward function, with total reward $R = T^2$ is increasing super-linearly in $T$. 
	From the variance computation of a constant reward function in Eq.~\eqref{eq:var_c},
	it follows that $\lim_{T\to \infty} {\rm Var}(v_{A,r_c}(T))= v_A(0)(1-v_A(0))$.
	
\end{proof}

From the analysis of a 
time-dependent P\'olya's urn model, we  know the variance satisfies the following formula (see proof  in Appendix \ref{sec:variance_proof} and also  \cite{Pem90}). 

\begin{lemma} 
	\label{lem:lem_variance}
	Let $e^{\theta_n} \triangleq  S(n)/S(n-1)$, then 
	\begin{eqnarray}
		\label{eq:variance}
	 {\rm Var}( v_{A,r}(T) ) =  \big(  v_{A,r}(0) - v_{A,r}(0)^2 \big)\Big( 1-  
	\frac{S(0)^2}{S(T)^2} \prod_{n=1}^{T} (2e^{\theta_n}-1) \Big)\;.
	\end{eqnarray}
	(Proof in Appendix \ref{sec:variance_proof})
\end{lemma}
 {Hence, although Definition \ref{def:equitability} applies to an arbitrary number of parties, 
Lemma \ref{lem:lem_variance} implies that it is sufficient to consider a single party's stake.
More precisely: 
\begin{remark} \label{rem:party}
If reward function $r:[T]\to\reals^+$ 
	over $T$  time steps is 
	{\em $\boldsymbol \varepsilon$-equitable} for vector $\boldsymbol \varepsilon = [\varepsilon_1, \ldots, \varepsilon_m]$ where $\varepsilon_i>0$, then $r$ is also $\tilde{\boldsymbol \varepsilon}$-equitable, where 
	$$
	\tilde{\boldsymbol \varepsilon} \triangleq \mathbf 1 \cdot \min_{i\in [m]}\varepsilon_i,
	$$
	with $\boldsymbol 1$ denoting the vector of all ones.
\end{remark}
As such, the remainder of this paper will study equitability from the perspective of a single (arbitrary) party $A$.
We will also describe reward functions as $\varepsilon$-equitable as shorthand for $\boldsymbol \varepsilon$-equitable, where ${\boldsymbol \varepsilon}=\mathbf 1 \cdot \varepsilon$.
}

Note that even if the total reward $R$ is fixed, 
equitability can differ dramatically across reward functions.  
In the example of Figure \ref{fig:beta}, the constant reward function is $0.5$-equitable.  
On the other hand, the geometric reward function of Eq.~\eqref{eq:reward_gi} 
has smaller chance of losing all its fractional stake (i.e.~$v_{A,r_g}(T)$ close to zero) 
or taking over the whole stake (i.e.~$v_{A,r_g}(T)$ close to one). 
It is $0.05$-equitable in this example. 

\section{Equitability under Honest Behavior} 
\label{sec:honest} 
In this section, we analyze the equitability of different block reward functions,
assuming that every party is honest, i.e.~follows protocol, and
the PoS system is  {\em closed}, so no stake is removed or added to the proposal stake pool 
over a fixed time period $T$. 
Each party's stake changes only because of the block rewards it earns and compounding effects.
We discuss the effects of strategic behavior in Section~\ref{sec:strategic}, and open systems in Section~\ref{sec:dynamic}.

The metric of equitability leads to a core optimization problem for PoS system designers: 
given a fixed total reward $R$ to be dispensed, 
how do we distribute it over the time $T$ to achieve the highest equitability? 
Perhaps surprisingly, we show that this optimization has a simple, closed-form solution. 

\begin{thm}
	\label{thm:honest_equitability}
	For all $R\in\reals^+$ and $T\in{\mathbb Z}^+$, 
	the geometric reward $\Rgi$ defined in \eqref{eq:reward_gi} is the most equitable 
	among  functions that dispense $R$ tokens over time $T${, jointly over all parties $A_i$, for $i\in [m]$.} 
\end{thm}

Intuitively, geometric rewards optimize equitability because they dispense small rewards in the beginning when the stake pool is small, 
so a single block reward cannot substantially change the stake distribution. 
The rewards subsequently grow proportionally to the size of the total stake pool,
 so the effect of a single block remains bounded throughout the time period. 
We emphasize that the geometric reward function 
only depends on $R$, $S(0)$, and $T$, and in particular 
does not depend on how the initial stake is distributed among the participating parties. 
Hence, it is universally most equitable for all parties in the system simultaneously. 


\subsection{Proof of Theorem \ref{thm:honest_equitability}}
{
Lemma \ref{lem:lem_variance} and Remark \ref{rem:party} imply that in order to show joint optimality over all parties, it is sufficient to show that for an arbitrary party $A$,}
	\begin{eqnarray}
		{\rm Var}\big(v_{A,\Rgi}(T)\big) \;\; \leq \;\;  {\rm Var}\big( v_{A,r}(T) \big) \;, 
	\end{eqnarray}
	for all $r\in\reals^T$ such that $\sum_{n=1}^T r(n)= R$ and $r(n)\geq0$ for all $n\in[T]$. 
To this end, we prove that $\Rgi$ is a unique optimal solution to the following optimization problem:
	\begin{eqnarray}
		\text{minimize}_{r\in \reals^T} && {\rm Var} (v_{A,r}(T)) \\ 
		\text{s.t.}&& \sum_{n\in[T]} r(n) = R \nonumber \\
		&& r(n)\geq 0\;,\; \forall n\in[T]. \nonumber
	\end{eqnarray} 
Using Lemma \ref{lem:lem_variance}, we have an explicit expression for ${\rm Var}( v_{A,r}(T) )$.
%
After some affine transformation and 
taking the logarithmic function of the objective, 
we get an equivalent optimization of 
	\begin{eqnarray}
		\text{maximize}_{\theta \in \reals^T} && \sum_{n=1}^T \log(2e^{\theta_n}-1) \\ 
		\text{s.t.}&& \sum_{n\in[T]} \theta_n =  \log (1+R) \nonumber\;, \\
		&& \theta_n \geq 0, \forall  n\in[T].
	\end{eqnarray} 
This is a concave maximization on a (rescaled) simplex. 
Writing out the KKT conditions with KKT multipliers $\lambda$ and $\{\lambda_n\}_{n=1}^T$, we get $\forall n \in [T]$: 
\begin{eqnarray}
	\frac{2e^{\theta_n}}{2e^{\theta_n}-1} - \lambda_n - \lambda &=& 0\\
	\lambda_n&\geq&0 \\
	\theta_n\lambda_n &=& 0
\end{eqnarray}
Among these solutions, 
we show that $\theta^*=((\log(1+R))/T) \, \ones$  is the unique optimal solution, 
where $\ones$ is a vector of all ones. 
 Consider a solution of the KKT conditions that is not $\theta^*$. 
 Then, we can strictly improve the objective by the following operation. 
 Let $i,j\in[T]$ denote two coordinates such that $\theta_i=0$ and $\theta_j\neq0$. 
 Then, we can create $\tilde\theta$ by mixing $\theta_i$ and $\theta_j$, such that 
 $\tilde\theta_n= \theta_n$ for all $n\neq i,j$ and 
 $\tilde\theta_i=\tilde\theta_j=(1/2)\theta_j$. 
We claim that $\tilde\theta$ achieves a smaller objective function as 
$\log(2e^{\theta_j}-1) < 2 \log(2e^{\theta_j/2}-1)$. 
This follows from Jensen's inequality and strict concavity of the objective function.
 Hence, $\theta^*$ is the only fixed point of the KKT conditions that cannot be improved upon. 
 
 In terms of the reward function, this translates into 
 $S(n)/S(n-1) = (1+R)^{1/T}$ and 
 $r(n)= (1+R)^{n/T} - (1+R)^{(n-1)/T}$.
 

 \subsection{Composition}
 \label{sec:multiple_interval}
The geometric reward function does not only optimize equitability for a single time interval. 
Consider a sequence $(T_1,R_1),  \ldots, (T_k, R_k)$ of checkpoints, 
where $T_i$ is increasing in $i$, and $R_i$ denotes the amount of reward to be disbursed between time $T_{i-1}+1$ and $T_i$ (inclusive).
These checkpoints could represent target inflation rates on a monthly or yearly basis, for instance. 
A natural question is how to choose a block reward function that optimizes equitability over all the checkpoints jointly.
The solution is to iteratively and independently apply geometric rewards over each time interval, giving a block reward function like the one shown in Figure \ref{fig:geom-reward}.

\begin{thm}
Consider a sequence of checkpoints $\{(T_i,R_i)\}_{i \in [k]}$. 
Let $\tilde R_j := \sum_{i=1}^j R_i$.
The most equitable reward function is 
\begin{equation}
r(n) = (1+\tilde R_{i-1})\left ( \left ( \frac{1+\tilde R_i}{1 + \tilde R_{i-1}} \right )^{\frac{n-T_{i-1}}{T_i-T_{i-1}}} -  \left ( \frac{1+\tilde R_i}{1 + \tilde R_{i-1}} \right )^{\frac{n-1-T_{i-1}}{T_i-T_{i-1}}} \right )
\label{eq:comp}
\end{equation}
for $n \in [T_{i-1}+1, T_{i}]$.
(Proof in Appendix \ref{app:composition})
\label{cor:composition}
\end{thm}
Notice that when there is only one checkpoint, Theorem \ref{cor:composition} simplifies to Theorem \ref{thm:honest_equitability}.
This result implies that checkpoints can be chosen \emph{adaptively}, i.e., they do not need to be fixed upfront to optimize equitability.
One practical concern with concatenating checkpoints in this manner is that the change in block rewards before and after a checkpoint can be dramatic, as seen in Figure \ref{fig:geom-reward}. 
This could cause other problems, such as proposers leaving the system.
Hence a PoS system need not choose its block reward function based on equitability alone; it could also consider smoothness and/or monotonicity constraints, for instance. 
We leave such investigations to future work.
Because of composition, we assume a single checkpoint for the remainder of this paper. 

\subsection{Equitability of Stake Pools}
\label{sec:pool} 

The participants have the freedom to form stake pools, 
as explored in \cite{schrijvers2016incentive,eyal2018majority,brunjes}. 
We show in this section that stake pool formation 
reduces the variances of the fractional stake of  all the members of the pool, and 
characterize exactly how much one gains. 
Consider a single party which owns $v_A(0)$ fraction of the stake at time $t=0$. 
We know from Lemma \ref{lem:lem_variance} that 
the variance at time $T$ is 
	\begin{eqnarray*}
	 {\rm Var}( v_{A,r}(T) ) =  \big(  v_{A}(0) - v_{A}(0)^2 \big)\Big( 1-  
	\frac{S(0)^2}{S(T)^2} \prod_{n=1}^{T} (2e^{\theta_n}-1) \Big)\;.
	\end{eqnarray*}
Consider a case  where 
the same party now participates in a stake pool, 
where the pool $P$ has $v_P(0)$ of the initial stake (including the contribution from party $A$), 
and every time the stake pool is awarded a reward for block proposal, the reward is evenly shared among the participants of the pool 
according to their stakes. 
The stake of party $A$ under this pooling is denoted by $v_{\tilde A}(T)$, 
and it follows from Lemma \ref{lem:lem_variance} immediately that 
	\begin{eqnarray}
	 {\rm Var}( v_{\tilde{A},r}(T) ) &=&  \Big( \frac{v_A(0)}{v_P(0)} \Big)^2\big(  v_{P}(0) - v_{P}(0)^2 \big)\Big( 1-  
	\frac{S(0)^2}{S(T)^2} \prod_{n=1}^{T} (2e^{\theta_n}-1) \Big) \nonumber \\ 
		&=& \frac{1-v_P(0)}{v_P(0)} \frac{v_A(0)}{1-v_A(0)} {\rm Var}(v_{A,r}(T))
	\;.
	\label{eq:pool}
	\end{eqnarray}
The party $A$'s variance reduces 
by a factor of $(v_P(0)/v_A(0))((1-v_A(0))/(1-v_P(0)))$ by joining a stake pool of size $v_P(0)$. 
For example, if everyone in the system form a single pool, then 
there is no randomness left and the variance is zero. 
Note that the variance is monotonically decreasing under stake pooling. 
In practice, 
stake pools can organically form as long as this gain in equitability exceeds the cost involved in forming such stake pools. 
Applying the definition \ref{def:equitability} to a single party $A$, 
equitability of a party improves by a factor of $(v_P(0)/v_A(0))((1-v_A(0))/(1-v_P(0)))$ in the sense that 
$\varepsilon$-equitable party $A$ will achieve 
$\varepsilon \frac{v_A(0)(1-v_P(0))}{v_P(0)(1-v_A(0))} $-equitability by forming a stake pool. 
Further, 
geometric reward function is still the most equitable reward function under the more general setting where 
the proposers are free to form stake pools. 
This follows from the fact that 
the effect of pooling is isolated from the effect of the choice of the reward function 
 in Eq.~\eqref{eq:pool},

\section{Practical Parameter Selection}
\label{sec:equitability_guideline} 

The equitability of a system is determined by 
four factors: the number of block proposals $T$, choice of reward function $r$, 
initial stake of a party $v_A(0)$, and the total reward $R$. 
We previously saw that geometric rewards optimize equitability over choices of the reward function; 
in this section, we study the dependence of equitability on $T$, $S(0)$, and $R$.
Recall that without loss of generality, we normalized the initial stake $S(0)$ to be one. 
For general choices of $S(0)$, the total reward  $R$ should be rescaled by $1/S(0)$. 
The evolution of the fractional stakes is exactly the same for 
one system with $S(0)=2$ and $R=200$ and 
another with $S(0)=1$ and $R=100$.
Although these parameters may be chosen according to external considerations (e.g.~interest rates, proposer incentives), 
we assume in this section that the system designer is free to choose the total reward $R$, 
either by setting the initial stake size $S(0)$ and/or 
by setting the total reward during $T$. 
We study how equitability trades off with the total reward  $R$ 
for different choices of   the reward function. 
Concretely, 
we consider a scenario where $r$ and $v_A(0)$ are fixed and $T$ is a large enough integer, and ask how many tokens we can dispense while 
maintaining a desired level of equitability $\varepsilon$. 

\subsection{Geometric reward function} For $\Rgi(n)$, we have $e^{\theta_n}=(1+R)^{1/T}$. 
It follows from Lemma \ref{lem:lem_variance} that 
\begin{eqnarray}
	\frac{{\rm Var}(v_{A,\Rgi}(T))}{ v_A(0)-v_A(0)^2 }  \;\; = \;\;  1 - \frac{(2(1+R)^{1/T} - 1)^T}{(1+R)^2}  \;, 
	\label{eq:var_g}
\end{eqnarray}
When $R$ is fixed and we increase $T$, we can distribute small amounts of rewards across $T$ and achieve vanishing variance. 
On the other hand, if $R$ increases much faster than $T$, 
then we are giving out increasing amounts of rewards per time slot and the uncertainty grows. 
This follows from the above variance formula, 
which we make precise in the following. 

\begin{remark}
	\label{rem:gi} 
	For a closed PoS system with a total reward $R(T)$ chosen as a function of $T$ 
	and a geometric reward function $\Rgi(n)=(1+R(T))^{n/T}-(1+R(T))^{(n-1)/T}$, 
	it is sufficient and necessary to set  
	\begin{eqnarray}
		R(T) \;\;=\;\; \left(\, \left(\,\frac{1}{1-\sqrt{\frac{\log(1/(1-\varepsilon))}{T}}}\, \right)^T -1\right) \,\big(1 + o(1)\big) \;,
	\end{eqnarray}  
	in order to ensure $\varepsilon$-equitability asymptotically, i.e. to ensure that
	\begin{eqnarray}
		 \lim_{T\to\infty} \frac{Var(v_{A,\Rgi}(T))}{v_A(0)(1-v_A(0))} = \varepsilon \;.
	\end{eqnarray}
\end{remark}

This follows from substituting the choice of $R(T)$ in the variance in Eq.~\eqref{eq:var_g}, 
which gives 
\begin{eqnarray}
	\lim_{T\to\infty} \frac{{\rm Var}(v_{A,\Rgi}(T))}{ v_A(0)-v_A(0)^2 }   &=& \lim_{T\to\infty}  1 - \Big( 1- \frac{\log(1/(1-\varepsilon))}{T} \Big)^T (1+o(1)).  \nonumber \\
	&=& \varepsilon\;, 
\end{eqnarray}
The limiting variance is monotonically non-decreasing in $R$ and non-increasing in $T$, as expected from our intuition. 
For example, if $R$ is fixed, one can have the initial stake $S(0)$ as small as $\exp(-\sqrt{T}/(\log T))$ and still 
achieve a vanishing variance.
As the geometric reward function achieves the smallest variance (Theorem \ref{thm:honest_equitability}), 
the above $R(T)$ is the largest reward that can be dispensed while achieving a desired 
normalized variance of $\varepsilon$ in time  $T$ (with initial stake of one). 
This scales as $ R(T)\simeq  (1+1/\sqrt{T})^{T} \simeq e^{\sqrt{T}}$.
We need more initial stake or less total reward, if we choose to use other reward functions.

\subsection{Constant reward function}
In comparison, consider the constant reward function of Eq.~\eqref{eq:reward_const}. 
As $e^{\theta_n}= (1+nR/T)/(1+(n-1)R/T)$, 
it follows from Lemma \ref{lem:lem_variance} that 
\begin{eqnarray}
	\frac{{\rm Var}(v_{A,\Rc}(T))}{v_A(0)-v_A(0)^2}  & = &  1 - \frac{1+R+\frac{R}{T}}{1+R+\frac{R}{T} + \frac{R^2}{T}}  \nonumber \\ 
	&=& \frac{R^2}{(T+R)(1+R)} \;. 
	\label{eq:var_c}
\end{eqnarray}
Again, this is monotonically non-decreasing in $R$ and non-increasing in $T$, as expected. 
The following  condition immediately follows from Eq.~\eqref{eq:var_c}. 

 \begin{remark}
	\label{rem:const} 
	For a closed PoS system with a total reward $R(T)$ chosen as a function of $T$ 
	and a constant reward function $\Rc(n)=R(T)/T$, 
	it is  sufficient and necessary to set  
	\begin{eqnarray}
		R(T) \;\;=\;\; \frac{\varepsilon\, T}{1-\varepsilon}\,  (1+o(1)) \;,
	\end{eqnarray}  
	in order to ensure $\varepsilon$-equitability asymptotically as $T$ grows. 
\end{remark}

By choosing  a constant reward function, 
the cost we pay is in the 
size of the total reward, which can now only increase as $O(T)$. 
Compared to $R(T)\simeq e^{\sqrt{T}}$ of the geometric reward, 
there is a significant gap. 
Similarly, in terms of how small initial stake can be with fixed total reward $R$, 
constant reward requires at least $S(0) \simeq R/T $. 
This trend gets even more extreme, for a decreasing reward function.

\subsection{Decreasing reward function}
Some cryptocurrencies use continuously \emph{decreasing} reward functions. 
For instance, Monero  dispenses block rewards as per 
\begin{eqnarray}
	\Rm(n) \;=\; \max \Big( c_0 \,,\, \frac{\lfloor (M - S(n) )  / c_1 \rfloor}{c_2}  \Big)\;,
	\label{eq:reward_Monero1}
\end{eqnarray}
for some constants $c_0$, $c_1$, $c_2$, and $M$.   
In practice, $c_0=0.6$, $c_1=2^{19}$, $c_2=10^{12}$ and $M=2^{64}$ \cite{miller2017empirical}.
Monero itself is not a PoS cryptocurrency, but if this decreasing reward were applied to the PoS setting,  it would have even higher variance than constant rewards. 
Consider a simpler choices of the constants such that 
\begin{eqnarray}
	\Rm(n) \;=\; \alpha(M-S(n-1) + S(0))\;,
	\label{eq:reward_Monero2}
\end{eqnarray}
for all $n\in[T]$, 
for a choice of $\alpha=1/T$ and $M=R(T)/(1-(1-\alpha)^T)$.
Recall that $S(n-1)-S(0)=\sum_{i=1}^{n-1} \Rm(i)$. 
As we assume $S(0)=1$, 
it follows after some calculations that 
$\Rm(n)= \alpha(1-\alpha)^{n-1}M$ 
and $S(n)=(1-(1-\alpha)^n)M + 1$. 
It follows from Lemma \ref{lem:lem_variance} that 
\begin{eqnarray}
	\frac{{\rm Var}(v_{A,\Rm}(T))} {(v_A(0)-v_A(0)^2)} \;=\; 1-\frac{S(0)^2}{S(T)^2}\prod_{n=1}^T \frac{1+M(1-(1-\alpha)^{n-1}(1-2\alpha))}{1+M(1-(1-\alpha)^{n-1})} \;. 
	\label{eq:var_M}
\end{eqnarray}

\begin{figure}[htbp]
\begin{minipage}{0.45\textwidth}
\begin{center}
\includegraphics[width=2.3in]{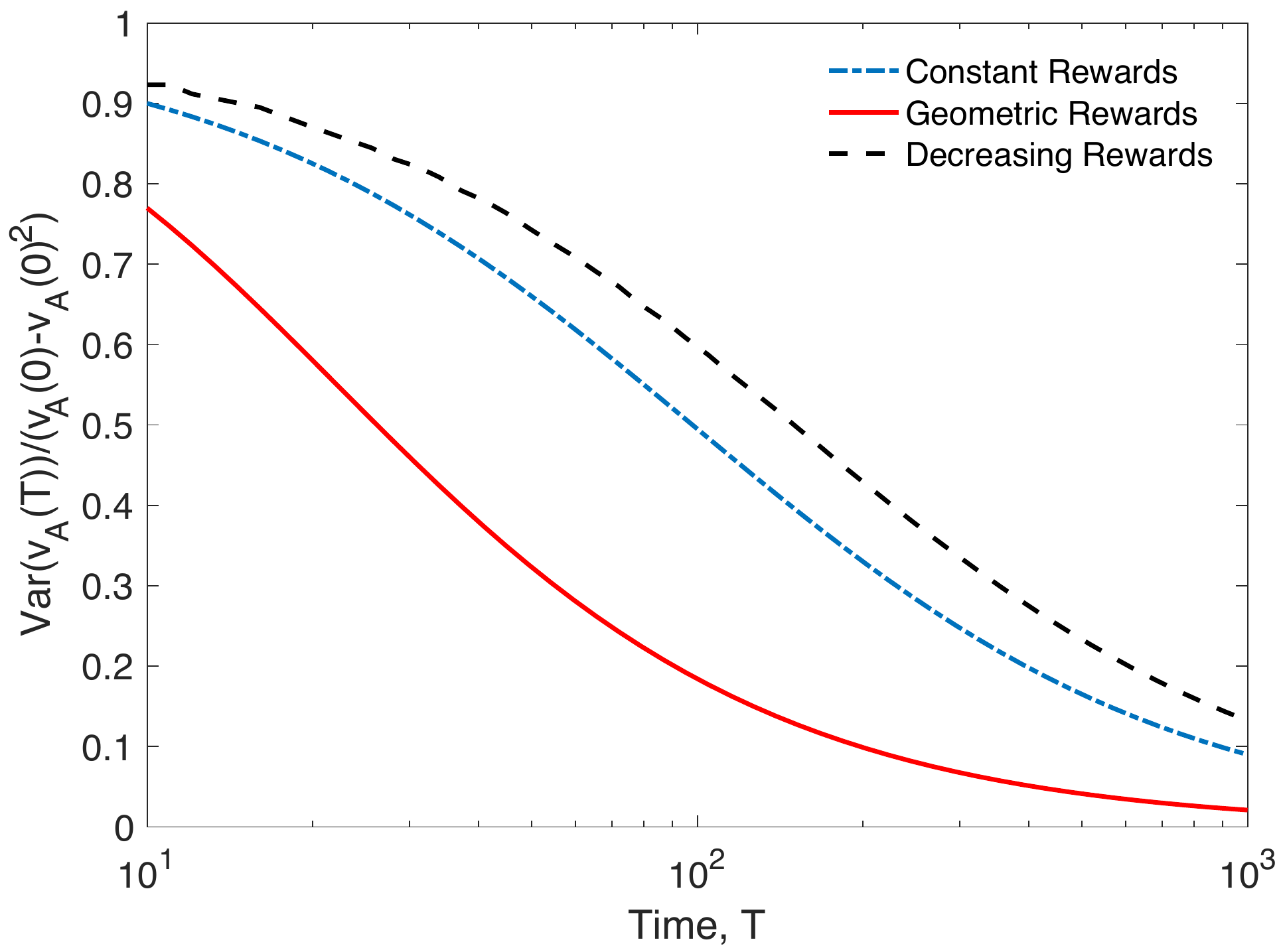}
\caption{Normalized variance after dispensing $R=10$ tokens over $T$ blocks, under different reward schemes.}
\label{fig:variance-reward}
\end{center}
\end{minipage}
~~
\begin{minipage}{0.45\textwidth}
\begin{center}
\includegraphics[width=2.3in]{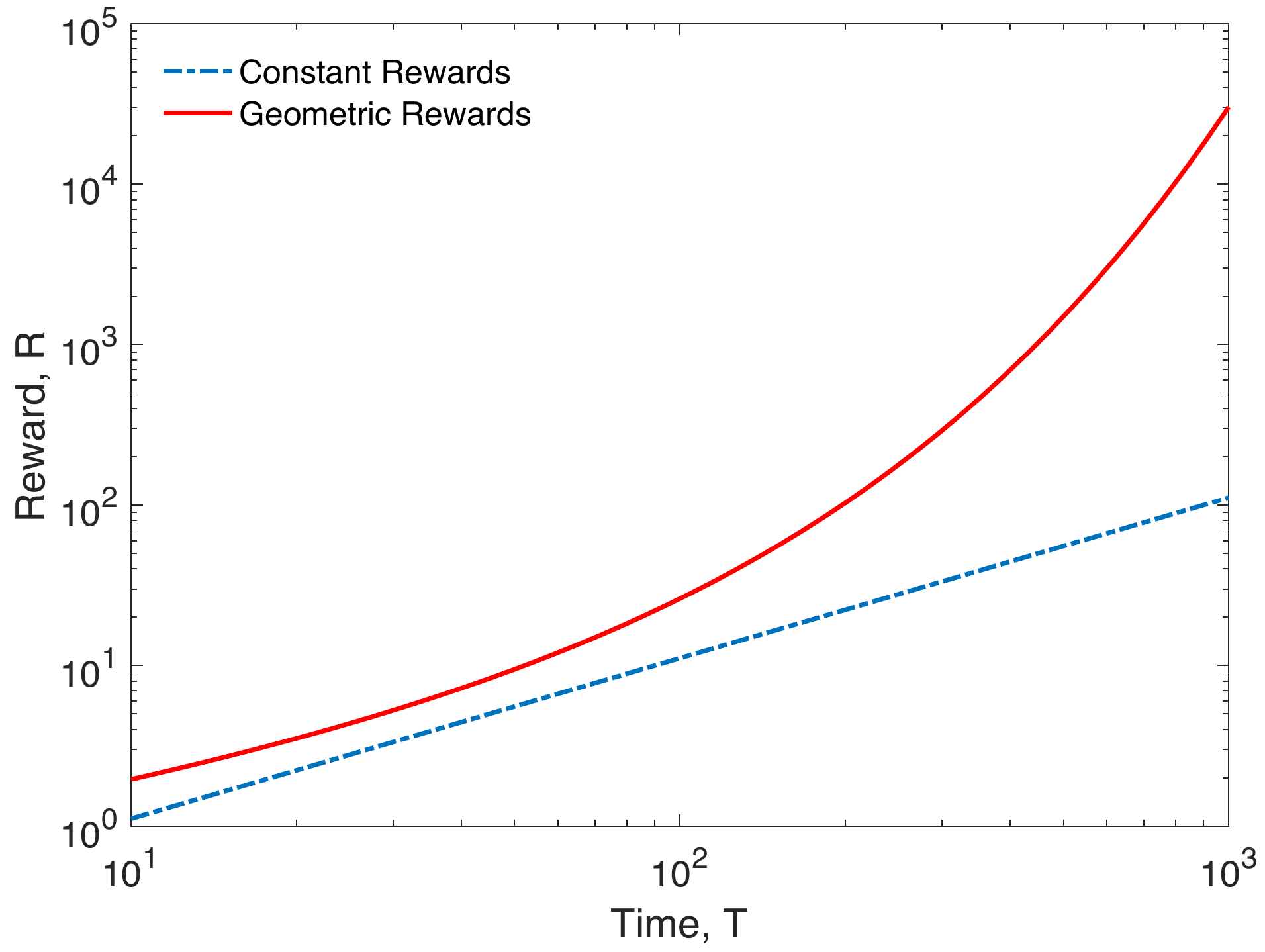}
\caption{Amount of reward that can be dispensed over $T$ blocks while guaranteeing a normalized variance of at most $\varepsilon=0.1$.}
\label{fig:reward_vs_time}
\end{center}
\end{minipage}
\end{figure}

\subsection{Comparison of Reward Functions}
For a choice of $S(0)=1$ and $R=10$,
Figure \ref{fig:variance-reward} illustrates the 
normalized variance for the three reward functions as a function of $T$, the number of blocks over which the reward is dispensed. 
As expected, variance decays with $T$ and geometric rewards exhibit the lowest normalized variance. 
Similarly, for a fixed desired (normalized) 
variance level of $\varepsilon=0.1$, Figure \ref{fig:reward_vs_time} shows how much the total reward can grow as a function of time $T$.
Notice that under constant rewards, the reward allocation grows linearly in $T$, whereas geometric rewards grow subexponentially fast while still satisfying the same equitability constraint. 

These observations add nuance to the ongoing conversation about how to initialize cryptocurrency  tokens that are \emph{not} considered securities from a regulatory perspective. 
In the 2018 class action lawsuit of Coffey vs. Ripple \cite{coffey}, one of the primary complaints against Ripple was the fact that ``all 100 billion of the XRP 
in existence were created out of thin air by Ripple Labs at its inception.''
Our results suggest that in a PoS system, a large initial stake pool can actually help to ensure equitability. 

\section{Strategic Behavior}
\label{sec:strategic}

In reality, proposers can be strategic to maximize their rewards. 
The most well-known strategic attack is \emph{selfish mining}, 
proposed by Eyal and Sirer in the context of proof-of-work \cite{eyal2018majority}, and extended in \cite{sapirshtein2016optimal,nayak2016stubborn}.
In selfish mining, adversarial miners who discover blocks do not immediately publish them; rather they build a private \emph{side-chain} of blocks. 
By eventually releasing a side chain that is longer than the main chain, the adversary can override blocks that were mined by the honest party. 
This has two effects: first, it gives the adversary a greater fraction of blocks in the main chain (and hence, block rewards) than they would get by mining honestly. 
Second, it forces honest parties to waste effort mining blocks that have a low chance of being accepted in the long term. 
Although selfish mining refers to a specific strategy that was designed for a PoW system, the concept of building side chains for increased profit can be applied to PoS systems as well. 
In this section, we show that such strategic attacks are exacerbated by the compounding effects of PoS. 
Contrary to the scenario where everyone behaves honestly, 
we empirically show that geometric reward functions do not mitigate the effects of compounding when strategic actors are present.

\subsection{Model}
\label{sec:strategic_model}
We restrict ourselves in this section to two parties: $A$, which is adversarial, and $H$, which is honest. 
Note that this is without loss of generality, as 
$H$ represent the collective set of multiple honest parties as their behavior is independent of how many parties are involved in $H$. 
The adversarial party $A$ can also represent the collective set of multiple adversarial parties, 
as having a single adversary $A$ is the worst case when all adversaries are colluding. 
Throughout this section, we use the terms 
{\em adversarial} and {\em strategic} interchangeably, to refer to the party that strategically deviates from honest behavior. 

Since $A$ does not always publish its blocks according to schedule, we distinguish the notion of a block slot (indexed by $n \in [T]$) and wall-clock time (indexed by $t \in [T]$).
It will still be the case that each \emph{block slot} $n$ has a single leader $W(n)$---in practice, this is determined by a distributed protocol---and a new block slot leader is elected at every tick of the wall clock (i.e., at a given time $t$, $W(n)$ is only defined for $n\leq t$). 
However, due to strategic behavior (i.e., the adversary can withhold its own blocks and override honest ones), it can happen that no block occupies slot $n$, even at time $t\geq n$; moreover, the occupancy of block slot $n$ can change over time. 
Thus, unlike our previous setting, if we wait $T$ time slots, the resulting chain may have fewer than $T$ blocks.
This is consistent with the adversarial model considered in PoS systems (like Ouroboros \cite{kiayias2017ouroboros}) that elect a single leader per block slot.
Other PoS systems, like PoSv3 \cite{posv3}, choose an independent leader to succeed each block; such a PoS model can lead to even worse attacks, which we do not consider in this work. 

\begin{figure}[htbp]
\begin{center}
\includegraphics[width=3.5in]{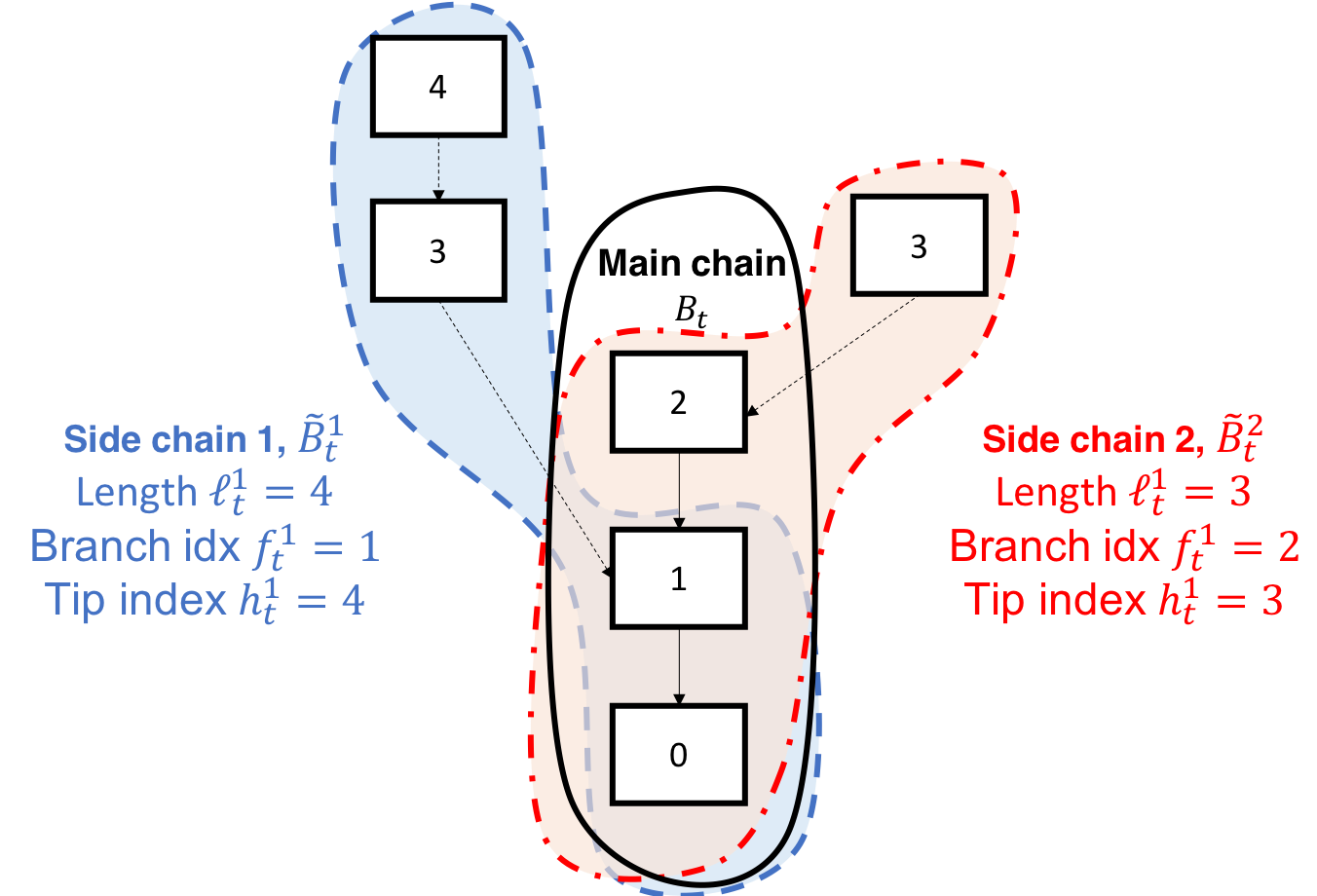}
\caption{In PoS, the adversary can keep arbitrarily many side chains at negligible  cost, and release (part of) a side chain whenever it chooses.}
\label{fig:side-chains}
\end{center}
\end{figure}

The honest party and the adversary have two different views of the blockchain, illustrated in Figure \ref{fig:side-chains}.
Both honest and adversarial parties see the  \emph{main chain} $B_t$; we let $B_t(n)$ denote the block (i.e., leader) of the $n$th slot, as perceived by the honest nodes at time $t$. 
If a block slot $n$ does not have an associated block at time $t$ (either because the $n$th block was withheld or overridden, or because $n>t$), we say that $B_t(n)=\emptyset$.
Notice that due to adversarial manipulations, it is possible for $B_t(n)=\emptyset$ and $B_{t-1}(n)\neq\emptyset$, and vice versa.

In addition to the main chain, the adversary maintains arbitrarily many private \emph{side chains}, $\tilde B_t^1, \ldots, \tilde B_t^s$, where $s$ denotes the number of side chains.
The blocks in each side chain must  respect the global leader sequence $W(n)$.
An adversary can choose at any time to publish a side chain, but we also assume that the adversary's attacks are \emph{covert}: it never publishes a side chain that conclusively proves that it is keeping side chains. 
For example, if the main chain contains a block $B$ created by the adversary for block slot $n$, the adversary will never publish a side chain containing block $\tilde B\neq B$, where $\tilde B$ is also associated with block slot $n$.

Each side chain ${\tilde B}_t^i$ with $i\in [s]$ overlaps with the honest chain in at least one block (the genesis block), and may diverge from the main chain after some $f_t^i \in \mathbb N_+$ (Figure \ref{fig:side-chains}).
That is, 
$$
f_t^i := \max \{n \in \mathbb N_+ \,:\, B_t(n) = \tilde B_t^i(n) \}.
$$
Different side chains can also share blocks; in reality, the union of side chains is a tree. 
However, for simplicity of notation, we consider each path from the genesis block to a leaf of this forest as a separate side chain, 
instead of considering side trees. 
We use $\ell_t$ and $\tilde \ell_t^i$ to denote the chain length of $B_t$ and $\tilde B_t^i$, respectively, at time $t$:
\begin{eqnarray*}
\ell_t =  | \{n \in [T] \,:\, B_t(n)\neq \emptyset\} | &\;, \;\; \text{ and } \;\;\;\;
\tilde \ell_t^i =  | \{n \in [T] \,:\, \tilde B_t^i(n)\neq \emptyset\} |,
\end{eqnarray*}
and we use the heights $h_t$ and $\tilde h_t^i$ to denote the block indices of the $\ell_t$th and $\tilde \ell_t^i$th blocks, respectively:
\begin{eqnarray*}
h_t =  \max \{n \in [T] \,:\, B_t(n)\neq \emptyset\}  &\;, \;\; \text{ and } \;\;\;\;
\tilde h_t^i =  \max \{n \in [T] \,:\, \tilde B_t^i(n)\neq \emptyset\}.
\end{eqnarray*}
If $f_t^i = h_t$, then  the adversary is building its $i$th side chain from the tip of the current main chain.

\subsubsection{State space.} 
The state space for the system consists of three pieces of data:
\begin{itemize}
\item The current time $t \in [T]$
\item The main chain $B_t$
\item The set of all side chains $\{\tilde B_t^i\}_{i \in [s]}$
\end{itemize}
Notice in particular that the set of side chains grows exponentially in $t$. 
In practice, most systems prevent the main chain from being overtaken by a longer side chain that branches more than $\Delta$ blocks prior to $h_t$; this is called a \emph{long-range attack}. 
Hence we can upper bound the size of the side chain set by imposing the condition that for all $i\in [s]$, $h_t - f_t^i \leq \Delta$.
Regardless, the size of the state space is considerably larger than it is in prior work on selfish mining in PoW \cite{sapirshtein2016optimal}, 
where the computational cost of creating a block forces the adversary to keep a single side chain. 

\subsubsection{Objective.}
The adversary $A$'s goal is to maximize its fraction of the total stake in the main chain by the end of the experiment, 
$$
v_{A}(t) \;\; = \;\; \frac{|\{n \in [T] \,:\, (W(n) = A) \land (B_T(n) \neq \emptyset) \}|}{\ell_T}.
$$
This objective is closely related to the metric of prior work \cite{sapirshtein2016optimal}, except for the finite time duration.

\subsubsection{Strategy space.}
The adversary has two primary mechanisms for achieving its objective: choosing where to append its blocks, and choosing when to release a side chain. 
If the honest party $H$ is elected at time $t$,  by the protocol, it always builds on the longest chain visible to it; since we assume small enough 
network latency, $H$ appends to block $B_{t-1}(h_{t-1})$. 
However,  if $A$ is elected at time $t$, $A$ can append to \emph{any} known block in $B_{t-1} \cup \{\tilde B_{t-1}^i\}_{i \in [s]}$. 
The system must allow such a behavior for robustness reasons: 
even an honest proposer may not have received a block $B_{t-1}(h_{t-1})$ or its predecessors due to network latency.

The adversary can also choose when to release blocks. 
In our model, $H$ always releases its block immediately when elected.
However, an adversarial proposer elected at time $t$ can choose to release its block at any time $ \geq t$; it can also choose not to release a given block. 
Late block announcements are also tolerated because of network latency; it is impossible to distinguish between a node that releases their blocks late and a node whose blocks arrive late because of a poor network connection.

 Notice that if $A$ is elected at time $t$ and chooses to withhold its block, the system advances to time $t+1$ without appending $A$'s block to the main chain.
 This means that the next proposer $W(t+1)$ is selected based on the stake ratios at time $t-1$. 
 So the adversary may have incurred a selfish mining gain from withholding its block, but it lost the opportunity to compound the $t^{\text{th}}$ block reward.
This tradeoff is the main difference between our analysis and prior work on selfish mining attacks in PoW systems.



Drawing from \cite{eyal2018majority,sapirshtein2016optimal}, at each time slot $t$, the adversary has three classes of actions available to it: match, override, and wait.
\begin{itemize}

\item The adversary  \textbf{matches} by choosing a side chain $\tilde B_t^i$ and releasing the first $h_t$ blocks. 
This means the released chain has the same height as the honest chain. 
In accordance with \cite{eyal2018majority,sapirshtein2016optimal}, we assume that after a match, the honest chain will choose to build on the adversarial chain with probability $\gamma$,  
which captures how connected the adversarial party is to the rest of the nodes. 

\item The adversary \textbf{overrides} by choosing a side chain $\tilde B_t^i$ and releasing the first $h = h_t+1$ blocks. 
The released chain becomes the new honest chain.

\item If the adversary chooses to \textbf{wait}, it does not publish anything, and continues to build on all of its side chains.

\end{itemize}

Unlike \cite{eyal2018majority,sapirshtein2016optimal}, we do not explicitly include an action wherein the adversary adopts the main chain. 
Because our model allows the adversary to keep an unbounded number of side chains, adopting the main chain is always a suboptimal strategy; 
it forces the adversary to throw away chains that could eventually overtake the main chain.
The primary nuance in the adversary's strategy is choosing \emph{when} to match or  override (rather than waiting), and \emph{which} side chain to choose. 
Identifying an optimal mining strategy through MDP solvers as in \cite{sapirshtein2016optimal} is computationally intractable 
due to the substantially larger state space in this PoS problem. 
Hence, in the following sections, we will discuss specific strategies that can increase the adversary's reward.

\subsection{Strategic selfish mining}
\label{sec:strategic_emp}

We show that 
the adversary can gain significantly by acting strategically, 
and this gain is 
exacerbated by the effect of compounding. 
Similar to selfish mining strategies originally introduced under PoW settings, 
an adversary can build side-chains to potentially take over the main chain. 
First critical difference is that 
an adversary can build arbitrarily many side chains branching from anywhere in the main-chain
without additional cost (other than the memory required to store those side-chains). 
In PoW, this is prevented by the computational power required to create each additional side-chain.
Secondly, the block rewards are also withheld for 
those adversarial blocks held aside to build side-chains. 
Under compounding, 
delaying the rewards of such side-chains costs the adversary in the following 
proposer elections, as the adversary is that much less likely to be elected a leader. 

An adversary needs to devise a strategy that balances 
the gain in keeping a long side-chain and potentially over taking a long main-chain, 
and the loss in those intermediate leader elections due to the withheld rewards. 
We propose a family of strategic schemes that we call 
{\em Match-Override-$k$} (MO-$k$).
Under MO-$k$ strategy, the adversary only keeps side-chains of length at most $k$. 

%

Concretely, the adversary uses the following strategy. 
Every time a new honest block is generated, 
this is appended to the main chain. 
At this point, the following actions are taken by the adversary. 
If there is a side-chain $\tilde B_t^i$ such that 
$h_t \neq f_t^i$ and $\tilde\ell_t^i \geq \ell_t $, 
then the adversary {\em matches} with the side-chain $\tilde B_t^i$ with the smallest $f_t^i$, 
 in which case the main chain length does not change,  
and the side chain $\tilde B_t^i$ remains, and all other side chains are discarded. 
Otherwise, 
if there is no such chain to match, 
then 
the main chain remains as is, 
the adversary {\em waits}, 
and any side-chain $\tilde B_t^i$ such that 
$\tilde\ell_t^i < \ell_t $ are discarded, as those side-chains are too short 
and have little chance of taking over the main chain. 
This action is known as {\em adopt} in the original selfish mining strategy. 

Every time a new adversarial block is generated, 
the adversary appends this block to every  side-chain she is managing currently. 
She also starts a new side-chain branching from the top of the main chain, 
if there is not a side-chain at the top already. 
At this point, one of the following actions are taken by the adversary. 
If there is only one side-chain $\tilde B_t^i$ and it satisfies both 
$h_t= f_t^i$ and $\tilde\ell_t^i \geq \ell_t +k$, then 
the adversary {\em overrides} with this $\tilde B_t^i$, 
in which case the main chain increments by one adversarial block, 
and the side chain $\tilde B_t^i$ remains.  
Otherwise, the main chain remains as is, and the adversary {\em waits}.

\begin{figure}
	\centering
	\includegraphics[width=.45\textwidth]{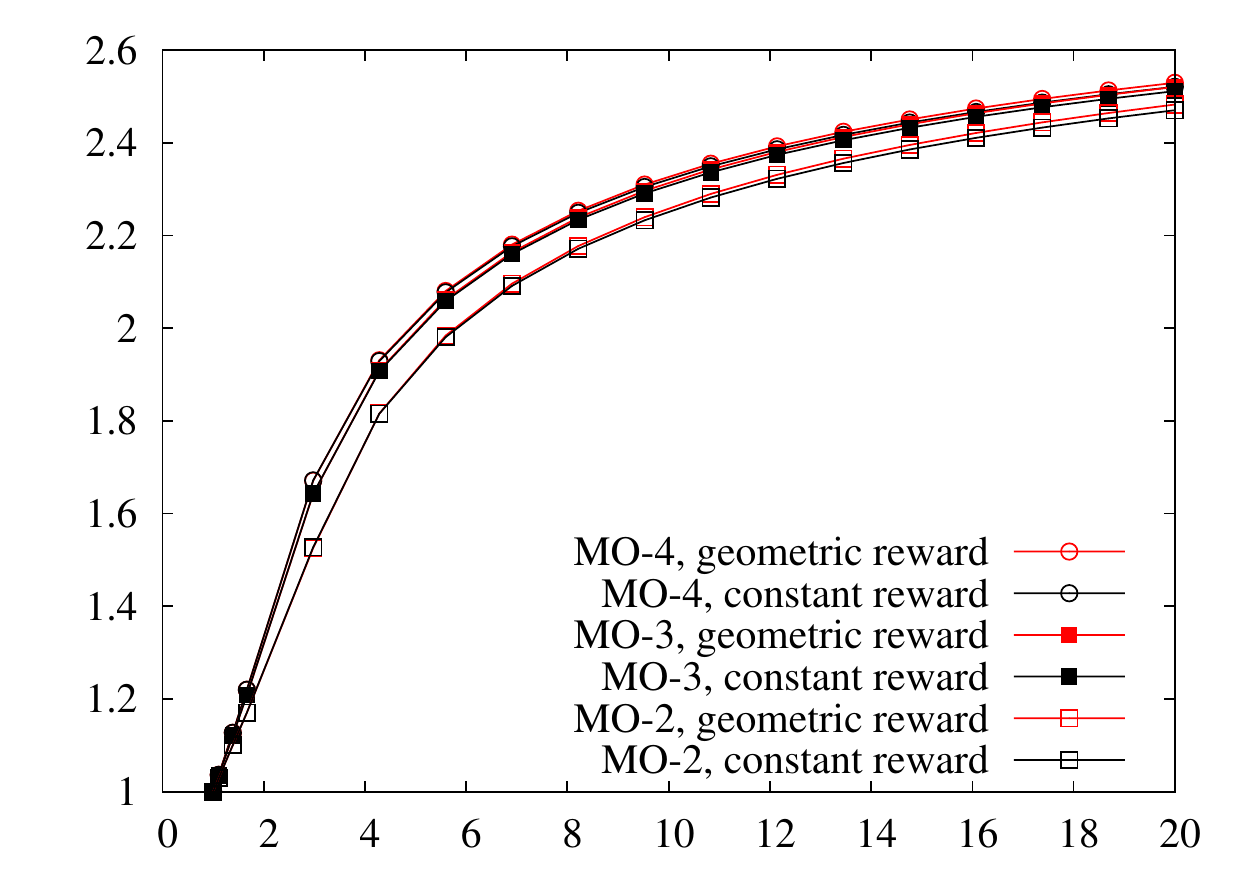}
	\put(-110,-10){$\frac{S(0)+R}{S(0)}$}
	\put(-230,90){$\frac{{\mathbb E}[v_A(T)]}{v_A(0)}$}
	\put(-110,145){$\gamma=1.0$}
	\hspace{0.3cm}
	\includegraphics[width=.45\textwidth]{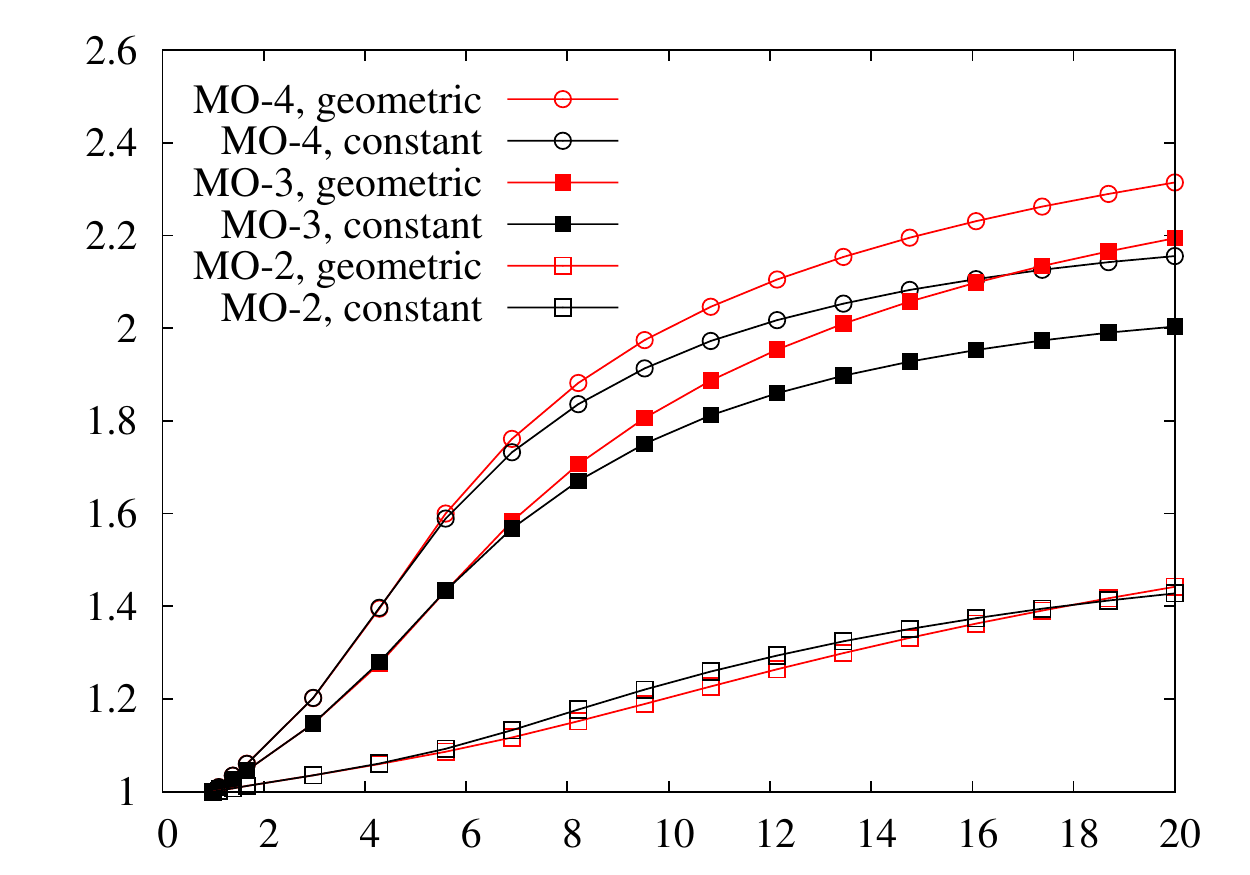}	
	\put(-110,-10){$\frac{S(0)+R}{S(0)}$}
	\put(-230,90){$\frac{{\mathbb E}[v_A(T)]}{v_A(0)}$}
	\put(-110,145){$\gamma=0.5$}
	\caption{Average fractional stake of an adversary can increase significantly 
	as the total reward $R$ increases. 
	We fix  initial fraction $v_A(0)=1/3$, $S(0)=1$, and $T=10,000$ time steps, and 
	show for two values of network connectivity of the adversary $\gamma\in\{0.5,1.0\}$ 
	defined in the strategy space subsection of Section \ref{sec:strategic_model} 
	and 
	varying total reward $R$.
	}
	\label{fig:strategic}
\end{figure}
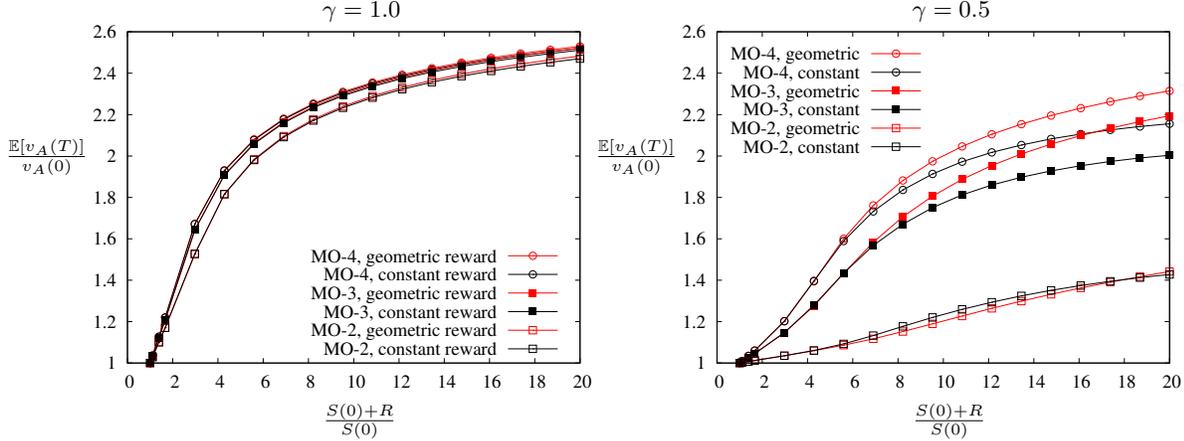

In Figure \ref{fig:strategic}, we can see that how much the adversary can gain in 
expected fractional stake, by using MO-$k$ strategies. 
As the total reward $R$ increases, the relative fractional stake approaches $3$, which is the maximum achievable value
as the expected fractional stake is normalized by $v_A(0)=1/3$. 
When the adversary is well connected to honest nodes, such that $\gamma=1.0$, such attacks are effective with  
small length side-chains, such as $k=3$ or $4$. Further, there is no distinguishable difference in the reward function used. 
On the other hand, when the adversary has equal chance of matching honest chains, such as $\gamma=0.5$, 
it is more effective to keep longer side-chains. 
Overall, the effect of strategic behavior is exacerbated by the compounding effects.

\subsection{Upper bound} 
\label{sec:strategic_am1}

We assume a constant reward function where a reward of $c$ is dispensed to a proposer whose block 
is appended to the main chain. 
We begin with an upper bound on $v_A(t)$, the fraction of stake that can be achieved by the adversary.

\paragraph{Always-Match-1 (AM-1):}
To show our upper bound, we analyze a random process called always-match-1 (AM-1). 
AM-1 is an urn process with state
$$
\mathbf X(t) = 
 \begin{bmatrix} 
X_A(t) \\
X_H(t)
\end{bmatrix}, 
\qquad
\mathbf X(0) = 
 \begin{bmatrix} 
S_A(0) \\
S_H(0)
\end{bmatrix},
$$
where as before, $S_A(t)$ denotes the number of tokens held by party $A$ at time $t$.
$X_H$ and $X_A$ can be thought of as the honest and adversarial stake, respectively;
compared to $S_A$ and $S_H$, they evolve under different dynamics, which are described below.
We let $v_A(t) := \frac{X_A(t)}{X_A(t)+X_H(t)}$ denote the fraction of the urn occupied by $X_A$ at time $t$.
At each tick of a discrete clock, the state is updated as follows:
\begin{eqnarray}
\mathbf X(t+1) = 
\begin{cases}
 \begin{bmatrix} 
X_A(t) \\
X_H(t)+c
\end{bmatrix} & \text{w.p. }  ~~ 1-v_A(t) \\ \\
 \begin{bmatrix} 
X_A(t)+c\\
\max\{X_H(t)-c,0\}
\end{bmatrix} & \text{w.p. } ~~ v_A(t).
\end{cases}
\end{eqnarray}
Intuitively, if the honest $X_H$ wins a given draw, then the honest pool gains $c$ unit of reward. 
If the adversarial pool $X_A$ instead wins a given draw, it negates $c$ honest units, and adds $c$ units to the adversarial pool.
The following theorem shows that AM-1 gives a {\em universal} upper bound on 
$v_A(t)$ under any arbitrary strategic behavior by the adversary. 
We refer to Appendix \ref{sec:am1_proof} for a proof. 

\begin{thm}
Under the constant reward function, 
for any adversarial strategy resulting in a stake fraction time series $v_A(t)$, the AM-1 random process $\tilde{v}_A(t)$ stochastically dominates $v_A(t)$, i.e.~
${\mathbb P}(v_A(t) \leq a ) \geq {\mathbb P}(\tilde{v}_A(t)\leq a)$ for all $a\in[0,1]$ and any $t\in{\mathbb Z}^+$.
	\label{thm:am1}
\end{thm}

\begin{figure}
	\centering
	\includegraphics[width=.45\textwidth]{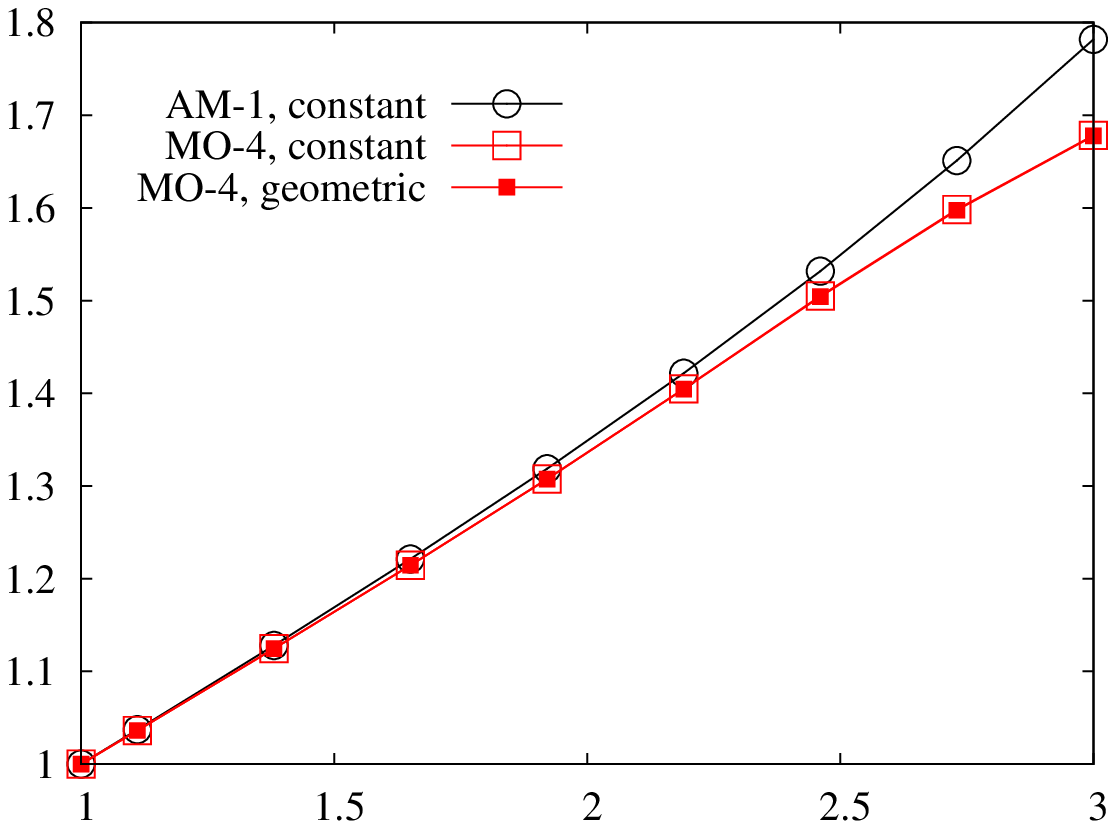}
	\put(-110,-10){$\frac{S(0)+R}{S(0)}$}
	\put(-230,90){$\frac{{\mathbb E}[v_A(T)]}{v_A(0)}$}
	\put(-110,145){$\gamma=1.0$}
	\hspace{0.2cm}
	\includegraphics[width=.45\textwidth]{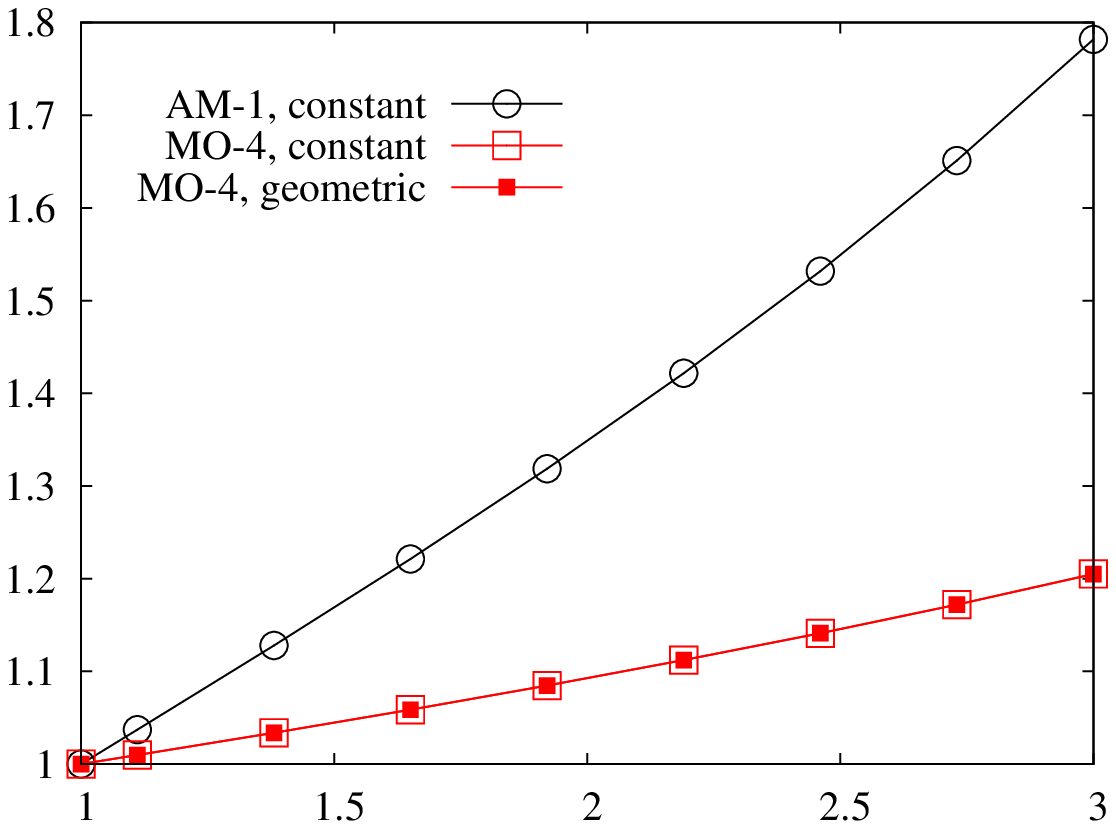}
	\put(-110,-10){$\frac{S(0)+R}{S(0)}$}
	\put(-230,90){$\frac{{\mathbb E}[v_A(T)]}{v_A(0)}$}
	\put(-110,145){$\gamma=0.5$}
	\caption{AM-1 urn process provides an upper bound on the compounding effect of any adversarial strategy. 
		We fix  initial fraction $v_A(0)=1/3$, $S(0)=1$, and $T=10,000$ time steps, and 
	show for two values of network connectivity of the adversary $\gamma\in\{0.5,1.0\}$ 
	defined in the strategy space subsection of Section \ref{sec:strategic_model} 
	and 
	varying total reward $R$.
	}
	\label{fig:am1}
\end{figure}

Figure \ref{fig:am1} (left) shows that for small values of the total reward 
$R\leq 2 S(0)$ and when adversaries are well connected to the honest nodes ($\gamma=1$), the AM-1 upper bound is quite close to 
an achievable strategy of MO-4. 
The right panel show 
that when the adversaries are less connected ($\gamma=0.5$), 
then the strategic behavior takes over less stake. 
We analyze an upper bound (inspired by AM-1), 
which reveals that a PoS system is less vulnerable against strategic attacks 
when initial stake $S(0)$ is larger.

\subsection{Analytical upper bound}
\label{sec:strategic_am2}

We introduce and analyze a new random process called always-match-2 (AM-2), 
which is an upper bound on AM-1, but has the merit that 
the expected fractional stake is tractable in a closed form. 

\paragraph{Always-Match-2 (AM-2):}
Similar to AM-1, AM-2 is an urn process with state
$$
\mathbf X(t) = 
 \begin{bmatrix} 
X_A(t) \\
X_H(t)
\end{bmatrix}, 
\qquad
\mathbf X(0) = 
 \begin{bmatrix} 
S_A(0) \\
S_H(0)
\end{bmatrix},
$$
where as before, $S_A(t)$ denotes the number of tokens held by party $A$ at time $t$.
At each tick of a discrete clock, the state is updated as follows:
\begin{eqnarray}
\mathbf X(t+1) = 
\begin{cases}
 \begin{bmatrix} 
X_A(t) \\
X_H(t)+c
\end{bmatrix} & \text{w.p. }  ~~ 1-v_A(t) \\ \\
 \begin{bmatrix} 
X_A(t)+2c\\
\max\{X_H(t)-c,0\}
\end{bmatrix} & \text{w.p. } ~~ v_A(t).
\end{cases}
\end{eqnarray}
The addition of $2c$ units of adversarial reward keeps the total change in urn size constant across time steps, which simplifies the analysis of this urn process. 
The following theorem shows that AM-2 gives an upper bound on 
the AM-1 process. 
We refer to Appendix \ref{sec:am2_proof} for a proof. 

\begin{thm} 
Under the constant reward function, the AM-2 process stochastically dominates the AM-1 process. 
	\label{thm:am2}
\end{thm} 

We are interested in 
how much an adversary can gain by acting strategically. 
The above theorem 
provides a tool for characterizing an upper bound on any strategies, 
by analyzing AM-2.
This is made formal in the following theorem. 
We refer to Appendix \ref{sec:strategic-upper_proof} for a proof. 

\begin{thm}
	\label{thm:strategic-upper}
	Let $v_A(t)$ denote the fractional stake of the adversary under the AM-2 process, 
	when the total initial stake is $S(0)$, 
	initial fractional stake of the adversary is $v_A(0)$, 
	and the total reward dispensed over time $T$ is $R=cT$. 
	If $R\leq S(0)(1-v_A(0))$, then 
	\begin{eqnarray}
		\E [v_A(T)]  \;\; = \;\; ( 1 + \eta ) \, v_A(0) \;,
		\label{eq:mean_am2}
	\end{eqnarray}
	where $\eta\triangleq R/(S(0)+c)$.
\end{thm}

Under the assumption that $R$ is less than the stake of the honest party to ensure that 
honest party's stake does not vanish to zero, 
the gain of an adversarial strategy over a honest strategy is bounded by 
$\E[v_A(T)] - \E[v_A(0)] \leq \eta \, v_A(0)$, 
where we used the fact that when everyone is honest the mean fractional stake remains 
$v_A(0)$ over all $t$. 
This implies that having a small initial stake $S(0)$ relative to the total reward $R$ makes the system 
vulnerable against adversarial strategies. 
This justifies the common practice of starting a PoS system with 
large initial stake. 

Further, this analysis allows us to 
quantify the price of compounding under adversarial strategy. 
When there is no compounding effect, 
either under a PoW system or because the rewards are not automatically appended to the stake, 
an upper bound on adversarial strategy we consider in this paper has been analyzed in \cite{sapirshtein2016optimal}. 
Translating the bound into the same notations as in Theorem \ref{thm:strategic-upper}, 
we get that when there is no compounding, an adversary's fractional stake is bounded by 
\begin{eqnarray*}
	v_A(T) \;\; \leq \;\; \Big( \,  1+\frac{v_A(0)\, \eta }{1+(1-v_A(0))\, \eta} \, \Big) \, v_A(0) + O\Big(\sqrt{\frac{\log T}{T}} \Big)  \;, 
\end{eqnarray*}
with high probability. 
For large enough $T$, with high probability. 
Compared to Eq.~\eqref{eq:mean_am2}, when $\eta$ is large, 
compounding allows the adversary's gain to grow linearly in $\eta$ 
whereas the adversary's gain is a constant in $\eta$ with no compounding. 
This shows that strategic parties can gain 
significantly over honest parties, under PoS systems with compounding effects.

\section{Discussion}
\label{sec:system}
There are three main issues that relate to actually building a chain-based PoS system with geometric rewards. 
The first is how to choose the relevant parameters $T$ and $R$, which has been discussed at length in  Section \ref{sec:equitability_guideline}.  
The second is how to deal with changing stake fractions that arise due to user-initiated transactions, e.g., selling their stake -- discussed below in Section \ref{sec:dynamic}).
The third discusses how to handle strategic behavior by block proposers in practice -- discussed here in Section \ref{sec:system-strategic}).

\subsection{Dynamic Proposer Stake} \label{sec:dynamic}
One challenge in the analysis of PoS systems is the fact that stake can move rapidly between parties, e.g. if nodes choose to sell their stake.
Computing the objective function in the optimization of equitability is tedious when accounting for the dynamic addition and removal of stake, 
and it is not clear that geometric rewards are robust to rapid stake transactions. 
However, in practice, PoS systems often restrict the timescale over which stake can be added or removed, precisely to add robustness. 
For example, Casper FFG constrains users to keep their stake in a validation pool for at least $4$ months in order to participate \cite{buterin2017casper,casper}. 

In our system, an analogous stability constraint would be to impose that stake ratios should not change during each time interval of  $T$ blocks. 
If this constraint is met, then geometric rewards can be recalculated at each block interval $T$ to account for dynamically changing stake pool. 
Theorem \ref{cor:composition} implies that this strategy optimizes the overall equitability of the reward scheme, even if the stake transactions are not known \emph{a priori}. 
Moreover, if we choose $T$ on the order of days as suggested in Section \ref{sec:equitability_guideline}, this constraint is relatively mild from a user's perspective.
It is important to note that users need not explicitly deposit their funds into a common pot in order to enforce the proposed stability constraints. 
This can be enforced implicitly by programming the selection mechanism to only consider stake that has been associated with the same public key for some minimum time interval.
Such a strategy has been suggested in several proposed PoS systems, including Ouroboros 
\cite{kiayias2017ouroboros}, Algorand \cite{micali2016algorand}, Casper \cite{buterin2017casper,casper}. 

\subsection{Control selfish mining} \label{sec:system-strategic}
Strategic behavior is a significant concern in PoW cryptocurrencies \cite{eyal2018majority,nayak2016stubborn,sapirshtein2016optimal}, and even more so in PoS systems. 
In Section \ref{sec:strategic} we demonstrate the efficacy of a strategic attack through which a rational user can artificially boost her proportion of the block rewards. 
In a sense, the results from Section \ref{sec:strategic} are negative.
Choosing a small reward (with respect to the initial fraction) at each time step does not fully solve the problem, and
there may be economic reasons to give out larger block rewards within a given time period. 

Ultimately, we expect that this problem cannot be solved solely by changing the block reward function. 
Rather, it may be more effective to control the \emph{effects} of strategic behavior than to identify a scheme under which  strategic behavior is equivalent to honest behavior. 
For instance, the proposer selection protocol could choose only proposers whose fraction of proposed blocks in the last $K$ blocks is commensurate with the proposer's stake (within some statistical error).
Such a policy would detect nodes who produce more than their fair share of blocks, and limit their ability to propose more blocks. 

\section{Conclusion}
\label{sec:conclusion}

In this work, we study the effects of compounding and 
the choice of block reward function on the concentration 
of wealth in PoS cryptocurrencies. 
We measure this concentration of wealth 
through a proposed metric called equitability, 
which captures the (normalized) variance of parties' stake distributions after a fixed epoch of $T$ blocks. 
We show that existing block reward functions (such as constant and decreasing rewards) 
have poor equitability. 
We introduce a new reward function, which we call geometric rewards, 
and prove that this is the most equitable block reward function. 
The negative effects of compounding, i.e.~the unfair distribution of wealth, 
can be further mitigated by choosing initial system parameters judiciously:
that is, by ensuring that the total block rewards disseminated in each epoch should be small compared to the initial stake pool size.

Several open questions remain.  
First, our results assume that proposers do not add or remove stake in the middle of an epoch. 
Such stake dynamics are likely to affect the optimality of geometric rewards and complicate the computation of equitability. 
Although we can disallow the addition or removal of stake on short timescales (e.g., a day), 
systems that choose epochs on the order of years will need to deal with dynamic stake pools. 

Another challenge, which we discuss in Section \ref{sec:system}, is that geometric rewards may not be desirable in practice because of the sharp changes in block rewards between epochs. 
A natural solution is to impose smoothness or monotonicity constraints on the class of reward functions. 
Solving such an optimization is an interesting direction for future work. 

Finally, a substantial open problem is that of protecting against strategic players. 
Although strategic players are not specific to PoS systems or compounding, we show here that geometric rewards alone do not protect against strategic players. 
Designing incentive-compatible consensus protocols for strategic players 
is a major question in blockchain systems.
Some papers that make progress on this front include Fruitchains \cite{fruitchains} and Ouroboros Hydra \cite{hydra}; 
both works propose reward and consensus mechanisms for which honest behavior is shown to be a $\delta$-approximate Nash equilibrium. 
As discussed in Section \ref{sec:related}, the algorithms of these papers may inherently improve equitability by spreading block rewards over multiple parties. 
Formally analyzing those protocols through the lens of equitability is another direction for future research.

\section{Acknowledgments}
The authors gratefully acknowledge support from the Distributed Technologies Research Foundation, the National Science Foundation under grant CCF 1705007, and the Army Research Office under grant W911NF1810332.

\bibliographystyle{acm}
\bibliography{references}

\newpage
\appendix
\section*{Appendix}
\section{Proof of Lemma \ref{lem:lem_variance}}
\label{sec:variance_proof}

	Let $e^{\theta_n} \triangleq  S(n)/S(n-1)$ and $r(n) = S(n+1)-S(n)$, then 
	\begin{eqnarray*}
		\E[v_{A,r}(n+1)^2|v_{A,r}(n)]  & =&  v_{A,r}(n) \Big(\frac{S(n) v_{A,r}(n) + r(n) }{S(n+1)}\Big)^2 + (1-v_{A,r}(n)) \Big(\frac{S(n)\, v_{A,r}(n) }{S(n+1)}\Big)^2  \\ 
			&=& \frac{ (S(n)^2 + 2 r(n)S(n))v_{A,r}(n)^2 +  r(n)^2 v_{A,r}(n) }{S(n+1)^2} \\ 
			&=& (2e^{-\theta_{n+1}} - e^{-2\theta_{n+1}} ) v_{A,r}(n)^2 + (e^{-\theta_{n+1}}-1)^2 v_{A,r}(n)\;. 
	\end{eqnarray*}
	It follows that 
	\begin{eqnarray*}
		\E[v_{A,r}(T)^2] - \E[v_{A,r}(T)] 
			&=& (2e^{\theta_{T}} - e^{2\theta_{T}} )\big( \E[v_{A,r}(T-1)^2] -\E[v_{A,r}(T-1)] \big)\\
			&=& \big( \E[v_{A,r}(0)^2] -\E[v_{A,r}(0)] \big) \prod_{n=1}^{T} (2e^{-\theta_{n}} - e^{-2\theta_{n}} )\;.
	\end{eqnarray*}
	Hence, 
	\begin{eqnarray*}
		 {\rm Var}( v_{A,r}(T) ) &=& \big( \E[v_{A,r}(0)] -  \E[v_{A,r}(0)^2] \big) \Big( 1- \prod_{n=1}^{T} (2e^{-\theta_{n}} - e^{-2\theta_{n}} )\Big) \\
		 &=&  \big(  v_{A,r}(0) - v_{A,r}(0)^2 \big)\Big( 1- \prod_{n=1}^{T} e^{-2\theta_n} \prod_{n=1}^{T} (2e^{\theta_n}-1) \Big) \\
		 &=&  \big(  v_{A,r}(0) - v_{A,r}(0)^2 \big)\Big( 1-  \frac{S(0)^2}{S(T)^2} \prod_{n=1}^{T} (2e^{\theta_n}-1) \Big)\;.
	\end{eqnarray*}

\section{Proof of Theorem \ref{cor:composition}}
\label{app:composition}
By the same logic as the proof of Theorem \ref{thm:honest_equitability}, the optimization problem of interest can be written as 
	\begin{eqnarray}
		\text{maximize}_{\theta \in \reals^{T_k}} && \sum_{n=1}^{T_k} \log(2e^{\theta_n}-1) \\ 
		\text{s.t.}&& \sum_{n=T_{i-1}+1}^{T_i} \theta_n =  \log \left ( \frac{1+\tilde R_i}{1+\tilde R_{i-1}} \right ) \;, \forall i \in [k] \nonumber \\
		&& \theta_n \geq 0, \forall  n\in[T_k]\;, \nonumber
	\end{eqnarray} 
where recall that $\theta_n = \frac{S(n)}{S(n-1)}$, and we define $T_0 := 0$.
Notice that this optimization problem is separable over the variables in different time intervals, so we can separately solve $k$ optimization problems, each of the form 
	\begin{eqnarray*}
		\text{maximize}_{\theta \in \reals^{T_i-T_{i-1}}} && \sum_{n=T_{i-1}+1}^{T_i} \log(2e^{\theta_n}-1) \\ 
		\text{s.t.}&& \sum_{n=T_{i-1}+1}^{T_i} \theta_n =  \log \left ( \frac{1+\tilde R_i}{1+\tilde R_{i-1}} \right )  \nonumber \\
		&& \theta_n \geq 0, \forall  n\in[T_{i-1}+1,T_i]
	\end{eqnarray*} 
for each $i \in [k]$.
Using the same KKT conditions as in Theorem \ref{thm:honest_equitability}, we get that $\theta_n^* = \frac{1}{T_i-T_{i-1}}\log (\frac{1+\tilde R_i}{1+\tilde R_{i-1}})$, which in turn implies that for $n \in [T_{i-1}+1, T_i]$, 
$$
S(n) = (1+\tilde R_{i-1})\left (\frac{1+\tilde R_i}{1+\tilde R_{i-1}} \right )^{(n-T_{i-1})/(T_i-T_{i-1})}
$$ and 
$$
r(n) = (1+\tilde R_{i-1})\left ( \left ( \frac{1+\tilde R_i}{1 + \tilde R_{i-1}} \right )^{\frac{n-T_{i-1}}{T_i-T_{i-1}}} -  \left ( \frac{1+\tilde R_i}{1 + \tilde R_{i-1}} \right )^{\frac{n-1-T_{i-1}}{T_i-T_{i-1}}} \right ).
$$

\section{Proof of Theorem \ref{thm:am1}}
\label{sec:am1_proof}

%

\begin{figure}[h]
	\centering
	\includegraphics[width=.65\textwidth]{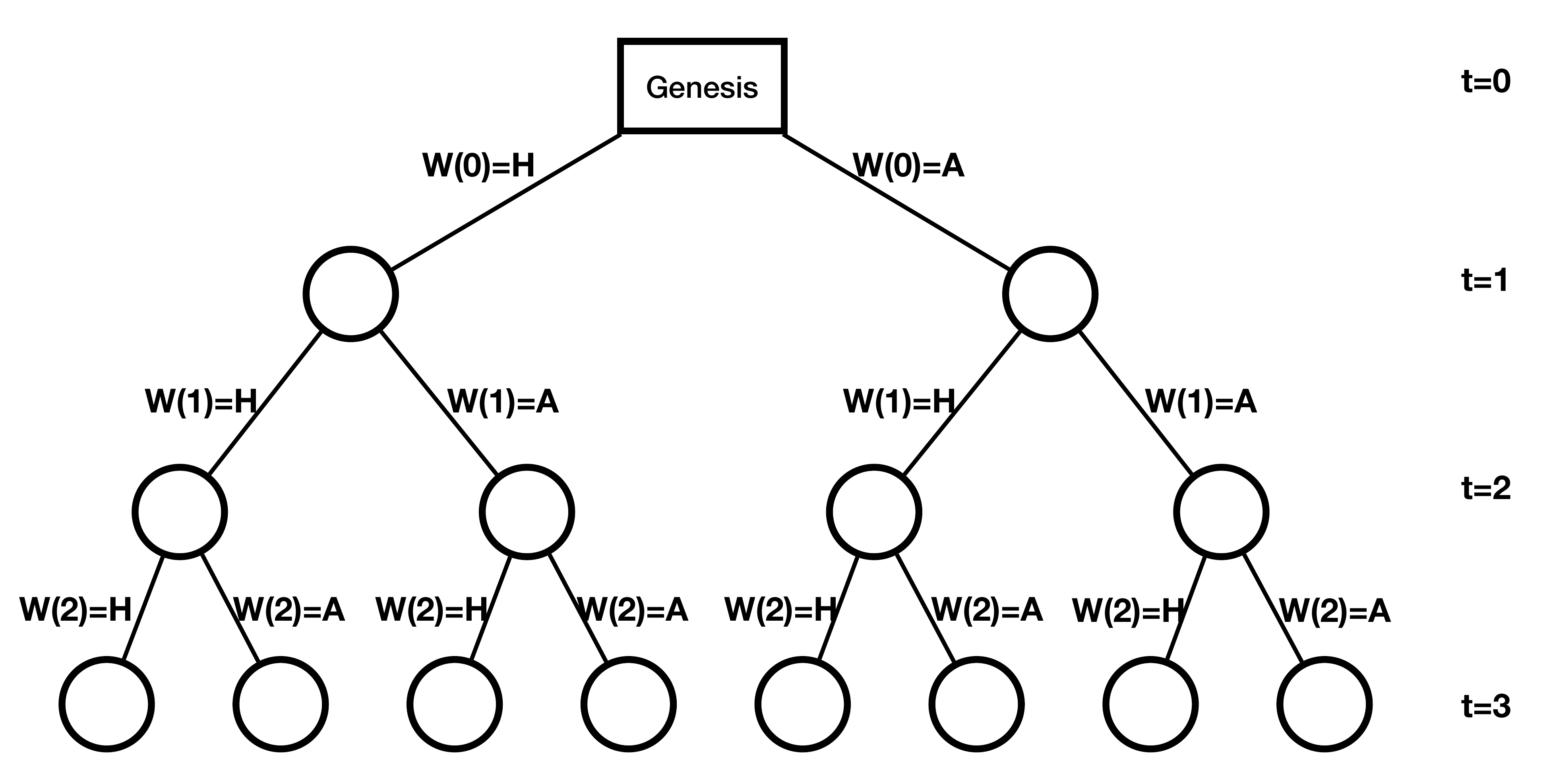}
\end{figure}

We first represent the standard P\`olya's urn process using a binary tree of the state evolution according to who won at each time step. 
Recall that the winner is assigned according to
\begin{eqnarray*}
	W(t) = \left\{ 
	\begin{array}{rl}
	H & \text{ with probability } 1-v_A(t) \;, \\
	A & \text{ with probability } v_A(t) \;,
	\end{array} \right. \;,
\end{eqnarray*}
which determines who gets the reward. 
We need the following notations for the proof. 
We denote the outcome of the random winner drawings as $W(0:2)=AAH$ if 
the winner at time $0$ is the adversary (meaning that the adversary was elected the leader and an adversarial block is generated), 
at time $t=1$ is the adversary, and at $t=2$ is the honest party. 
Under  this event, we denote the factional stake of the adversary by $\tx(AAH)$, 
and the total stake by $S(AAH)$. 
We use the notion of the 
standard P\`olya's urn process to denote the process with constant reward $c$ at each time.

A strategic behavior consists of union of the following actions. 
When elected a leader at a certain time $t$, say $t=1$ in the figure below, 
the adversary may decide to withhold its currently generated block and also the reward.  
This withheld reward (and the block) is awarded when 
the adversary either matches or overrides (based on the instance of the future winner elections). 
If {\em matched}, the reclaimed block also takes away one of the honest blocks (and the corresponding reward). 
If {\em overridden}, the reclaimed block may or may not take away one of the honest blocks.  

We represent the strategy of the adversary on a {\em single} withheld block (generated at time $t=1$), 
using the following binary decision tree. 
We consider a binary decision tree of height $T$, as follows (e.g.~$T=4$). 
We can encode any strategy of the adversary on when to reclaim the withheld reward 
on the binary tree. We are hiding the part of the tree branching from $W(0)=H$ as it is not affected by the current adversary we are considering. 

\begin{figure}[h]
	\centering
	\includegraphics[width=.75\textwidth]{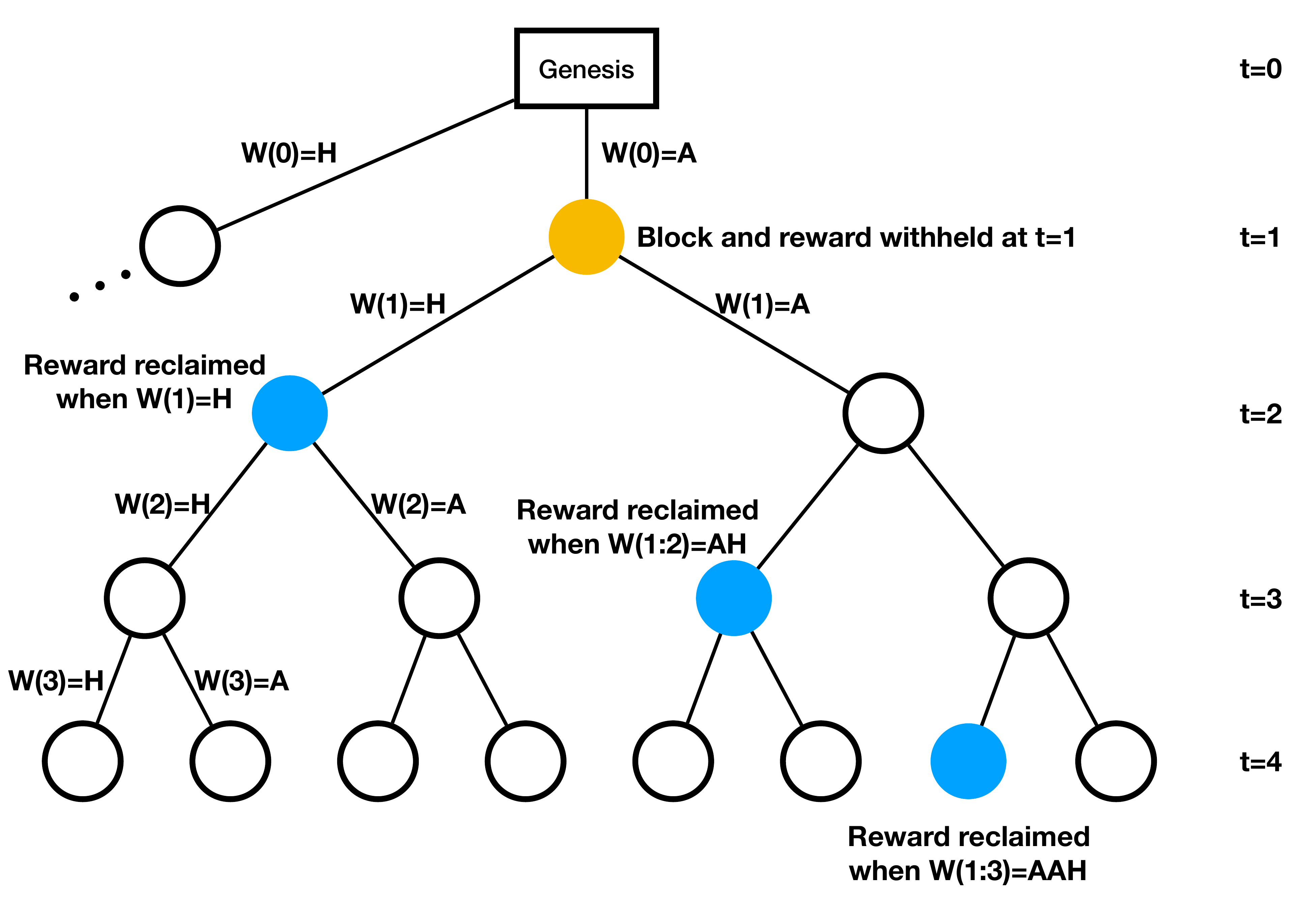}
\end{figure}

For example, 
the adversary withholds a block if he is elected a leader at time $t=1$, which is encoded as the orange node. 
The adversary might choose to 
reclaim the reward (by publishing the side-chain that includes the withheld block of interest) 
if the next winner is an honest one ($W(1)=H$), 
 if next two winners are adversarial and honest in that order ($W(1:2)=AH$), 
and if next three winners are adversarial, adversarial, and honest in that order ($W(1:3)=AAH$). 
These are encoded on a binary tree as shown above in blue nodes. 
Any binary tree where a single path from an orange node (block withheld) to a leaf only contains 
a single blue node (block reclaimed) is a valid strategy, as a withheld reward can only be claimed once. 
We do not explicitly encode wether a honest block is taken away when adversarial block is reclaimed, 
as it does not change the proof as we will show. 

In the above example, the orange node at the event $W(0)=A$
encodes the strategy that a unit $c$ reward is withheld if the adversary wins at time $T=0$. 
Hence, the resulting stake at that node is $\tx(A) = \tX(0)$ as no reward is claimed, 
and $S(A)=S(0)$.    
the blue node at the event $AH$ denotes that the adversary 
reclaims the withheld reward if the next winner is an honest party. 
Under this event of $AH$, 
the resulting stake at that node $AH$ 
is $\tx(AH) = \tX(A)+(c/S(A))$ as $c$ unit reward is given to the adversary and 
$c$ unit reward is taken from the honest party, and 
and the total stake $S(AH)=S(A)$ remains unchanged.     

The next lemma provides a set of operations on the colored binary tree we can perform, 
in order to turn it into a more stochastically dominant process. 
We give a proof in Appendix \ref{sec:ruleA_proof}.

\begin{lemma} Given a representation of a random process with an adversarial strategy as a colored binary tree, 
	the following operation results in a new random process that stochastically dominates the old one:
	\label{lem:ruleA}
	\begin{itemize}
		\item[A.1.] convert a white leaf node to a blue leaf node; 
		\item[A.2.] convert two blue leaf nodes who are siblings into one blue parent node  with two white offsprings; 
		\item[A.3.] convert two blue sibling nodes into one blue parent node with two white offsprings; and 
		\item[A.4.] convert two blue offsprings of an orange node into one parent node that is blue and orange with two white offsprings.
	\end{itemize}
\end{lemma}
Note that a blue and orange node denotes the combination of a orange node and a blue node, where 
one unit reward given to the adversary and one unit reward taken away from the honest party. 
Applying the above operations in the order of $A.1$, $A.2$, $A.3$, and $A.4$, 
we have the following random urn process that stochastically dominates any adversarial behavior with a {\em single} 
reward withheld. 
\begin{figure}[h]
	\centering
	\includegraphics[width=.9\textwidth]{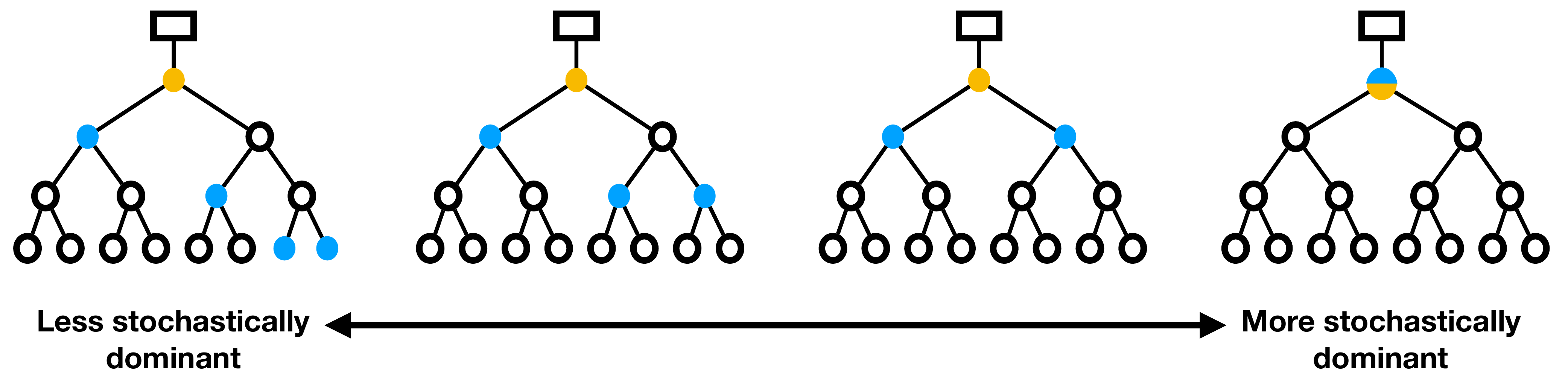}
\end{figure}

From the preservation of stochastic dominance by the standard P\`olya's urn process 
(as shown in Lemma \ref{lem:dom_polya}), we can now 
convert a white node in which an adversary is a winner into a blue and orange node, from top to bottom. 
The resulting process is exactly the AM-1 process, finishing the proof of stochastic dominance when only a {\em single} reward is withheld at time $t=1$. 
\begin{figure}[h]
	\centering
	\includegraphics[width=.9\textwidth]{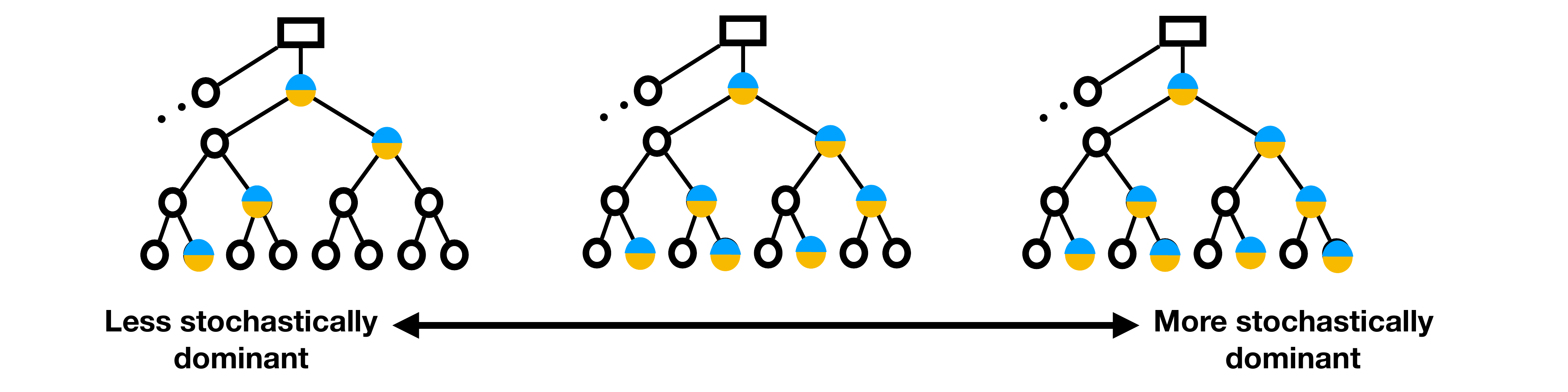}
\end{figure}

When a single reward is withheld at time $t>1$, we need a preservation of 
stochastic dominance for AM-1. The following lemma justifies conversion of a white node into a blue and orange node 
when the descendent nodes follow AM-1. We provide a proof in Appendix \ref{sec:dom_am1_proof}. 

\begin{lemma}[Preservation of stochastic dominance of AM-1 process]
	\label{lem:dom_am1}
	Consider two AM-1  processes with 
	the same initial total stake $S(0)$. 
	One process has a random initial fractional stake $v_A(0)$, which is stochastically dominated 
	by that of the other process $v'_A(0)$. 
	Then, the final fractional stake preserves the dominance, i.e.~$v_A(T)$ is stochastically dominated by $v'_A(T)$. 
\end{lemma}

In general, a strategic behavior consists  of multiple rewards withheld at multiple nodes in the binary tree. 
Each withheld reward will have some strategy for being reclaimed in the future. 
Lemmas \ref{lem:dom_polya} and \ref{lem:dom_am1} ensure that the above argument for 
converting such a strategy into AM-1 process still holds when multiple rewards are withheld. 
This finishes the proof of the claim that AM-1 stochastically dominates 
any adversarial strategy.

\section{Proof of Theorem \ref{thm:am2}}
\label{sec:am2_proof}

The fact that AM-2 further stochastically dominates AM-1 follows immediately from Lemma 
\ref{lem:dom_am1} and converting a binary tree representation of  AM-1 to that of AM-2 
from top to bottom. We omit this part of the proof as it is straightforward. 

\section{Proof of Lemma \ref{lem:ruleA}}
\label{sec:ruleA_proof}

We use the following example of a adversarial strategy for illustration of the proof: 
\begin{figure}[h]
	\centering
	\includegraphics[width=.75\textwidth]{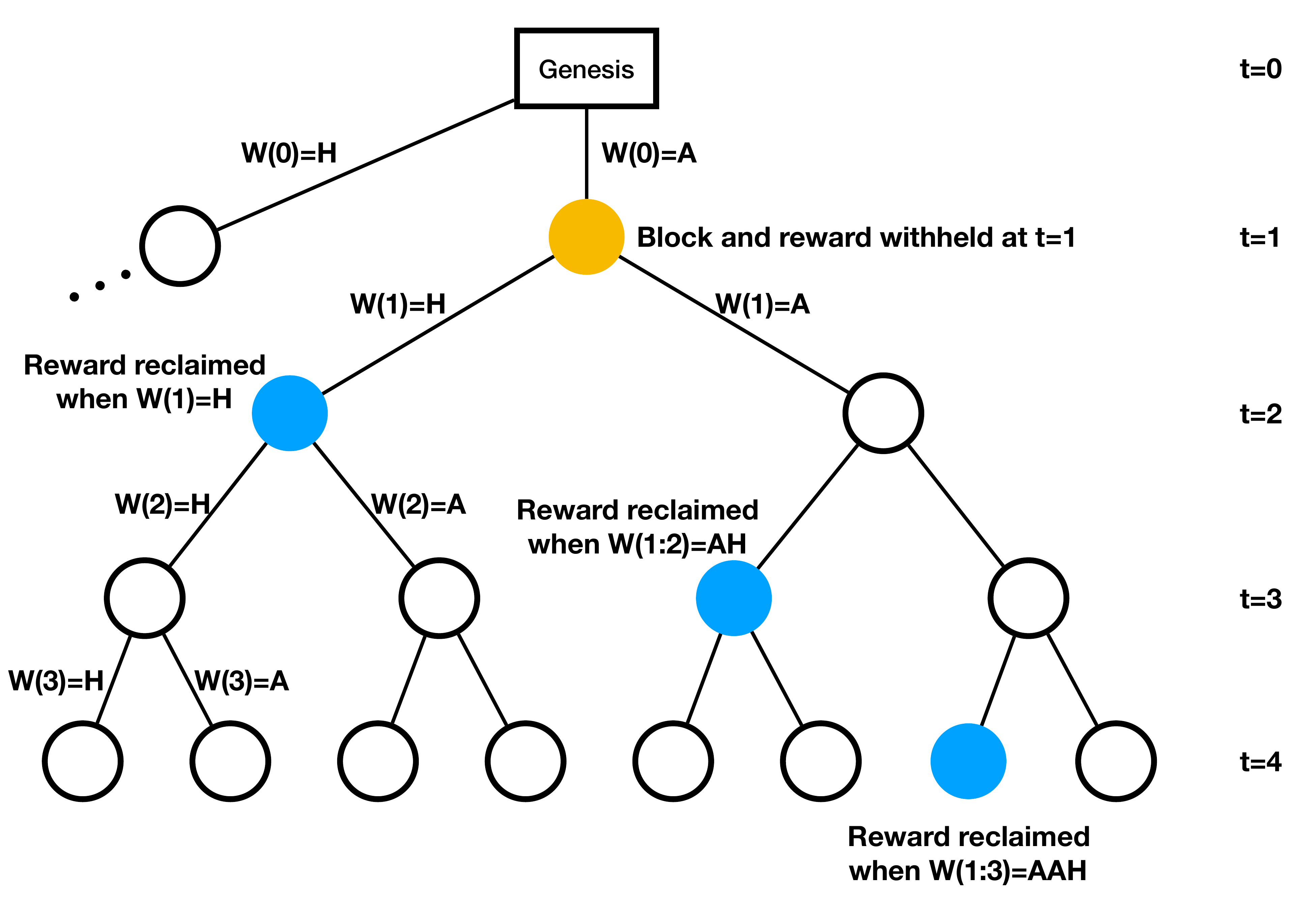}
\end{figure}

\noindent{\bf $A.1.$ Convert a white leaf node to a blue leaf node.}
As this change only affects one sample path, 
only one instance is affected, say $T=4$ and $W(0:T-1)=AAAA$.  
The probability of this outcome does not change, 
but only the fractional stake corresponding to this outcome changes from 
$\tx(AAAA)$ to $\tx'(AAAA)=\tx(AAAA) + (c/ S(AAAA))$. 
As $c/ S(AAAA)>0$, the process after the changing the white leaf node into a blue one is strictly 
stochastically dominant. 
After this conversion, the node $AAAA$ is now blue, in which case we apply the next operation. 

\bigskip
\noindent{\bf $A.2.$ Convert two blue leaf nodes who are siblings into one blue parent node with two white offsprings.} 
This change affects two sample paths, say $T=4$, $W(0:T-1)=AAAA$ and $W(0:T-1)=AAAH$.  	
The fractional stakes do not change, but the probabilities do. 
$\prob(AAAH|AAA)=1-\tx(AAA)$ changes to $\prob'(AAAH|AAA)=1-\tx(AAA)-(c/S(AAA))$. 
$\prob(AAAA|AAA)=\tx(AAA)$ changes to $\prob'(AAAA|AAA)=\tx(AAA)+(c/S(AAA))$. 
Given that $\tx(AAAH)\leq \tx(AAAA)$ and $c/S(AAA)>0$, 
the resulting process is stochastically dominant.  
After this conversion, the node $AAA$ is now blue, in which case we apply the next operation. 

\bigskip
\noindent{\bf $A.3.$ Convert two blue sibling nodes into one blue parent node with two white offsprings.} 
This change affects many sample paths, 
all descendants of a single node (parent of the two blue nodes of interest), 
say node $AA$, and 
node $AAA$ and node $AAH$ are blue nodes. 
We know from rule $A.2.$ that 
conditioned on the event $W(0:1)=AA$, 
 $\tx'(2)$ stochastically dominates $\tx(2)$. 
 The rest of the process follows the standard P\`olya's urn process. 
 Hence, the following lemma implies the desired claim. 
We provide a proof of this lemma in Appendix \ref{sec:dom_polya_proof}. 
\begin{lemma}[Preservation of stochastic dominance of the standard P\`olya's urn process]
	\label{lem:dom_polya}
	Consider two standard P\`olya's urn processes with 
	the same initial total stake $S(0)$ and the same constant reward $c$. 
	One process has a random initial fractional stake $v_A(0)$, which is stochastically dominated 
	by that of the other process $v'_A(0)$. 
	Then, the final fractional stake preserves the dominance, i.e.~$v_A(T)$ is stochastically dominated by $v'_A(T)$. 
\end{lemma}
After this conversion, the node $AA$ is now blue, in which case we apply the next operation.

\bigskip
\noindent{\bf $A.4.$ Convert two blue offsprings of an orange node into one parent node that is blue and orange with two white offsprings.}
Note that the stakes at time $t=2$ remain unchanged by the conversion, i.e.~$\tx(AA)=\tx'(AA)$, and $\tx(AH)=\tx'(AH)$. 
Only the corresponding probability of events change. 
Node $A$ is orange with, say $\tx(A)=v_A(t=0)$, as the reward is withheld at time $t=1$.  
Hence, $\prob(AA|A)=v_A(0)$, whereas 
$\prob'(AA|A)=v_A(0)+(c/S(A))$, after the conversion.
It follows from the fact that $c/S(A)>0$ and $\tX(AA)>\tX(AH)$ (and also Lemma \ref{lem:dom_polya}) that 
the conversion results in a process that is stochastically dominant.

\section{Proof of Lemma \ref{lem:dom_polya}}	
\label{sec:dom_polya_proof}

We prove it by a recursion. 
Consider the following representation of the standard P\`olya's urn process. 
$v_A(t)$ denotes the fractional stake of party $A$ at time $t$, that starts as $v_A(0)$. 
$S(t)= S(0)+c\,t$ denotes the total stake. 
First, we claim that if $v_A(0) < v'_A(0)$ deterministically, 
then $v_A(1) \stackrel{{\cal D}}{\leq} v'_A(1)$. 
This follows from the fact that 
\begin{eqnarray}
v_A(1) = \left\{ 
	\begin{array}{rl}
		\frac{v_A(0)S(0) + c}{S(0)+c} & \text{ w.~p. } v_A(0)\;, \\
		\frac{v_A(0)S(0)}{S(0)+c} & \text{ w.~p. } 1-v_A(0)\;,
	\end{array}
	\right. \;, \;\; \text{ and } \;\;\;\; 
v'_A(1) = \left\{ 
	\begin{array}{rl}
		\frac{v'_A(0)S(0) + c}{S(0)+c} & \text{ w.~p. } v'_A(0)\;, \\
		\frac{v'_A(0)S(0)}{S(0)+c} & \text{ w.~p. } 1-v'_A(0)\;,
	\end{array}
	\right. \;. 
\end{eqnarray}
For these two valued discrete random variable, 
not only are those two values both larger for the latter process, 
but also the probability mass for the larger of those two values are also 
higher for the latter process. 
Hence,  $v_A(1) \stackrel{{\cal D}}{\leq} v'_A(1)$. 
Note that assuming stochastic dominance of 
 $v_A(0) \stackrel{{\cal D}}{\leq} v'_A(0)$ leads to the same conclusion. 
 Hence, we can recursively apply the above result to prove the desired lemma.

\section{Proof of Lemma \ref{lem:dom_am1}}
\label{sec:dom_am1_proof}

Consider the following representation of the AM-1  process. 
Let $v_A(t)$ denote the fractional stake of party $A$ at time $t$, that starts as $v_A(0)$. 
Let $S(t)$ denote the total stake at time $t$, that starts with $S(0)$. 
First, we claim that if $v_A(0) < v'_A(0)$ deterministically, 
then $v_A(1) \stackrel{{\cal D}}{\leq} v'_A(1)$. 
This follows from the fact that 
\begin{eqnarray}
v_A(1) = \left\{ 
	\begin{array}{rl}
		\frac{v_A(0)S(0) + c}{S(0)} & \text{ w.~p. } v_A(0)\;, \\
		\frac{v_A(0)S(0)}{S(0)+c} & \text{ w.~p. } 1-v_A(0)\;,
	\end{array}
	\right. \;, \;\; \text{ and } \;\;\;\; 
v'_A(1) = \left\{ 
	\begin{array}{rl}
		\frac{v'_A(0)S(0) + c}{S(0)} & \text{ w.~p. } v'_A(0)\;, \\
		\frac{v'_A(0)S(0)}{S(0)+c} & \text{ w.~p. } 1-v'_A(0)\;,
	\end{array}
	\right. \;. 
\end{eqnarray}
The rest of the proof follows similarly as in the proof of  Lemma \ref{lem:dom_polya} in 
Section \ref{sec:dom_polya_proof}. 

\section{Proof of Theorem \ref{thm:strategic-upper}}
\label{sec:strategic-upper_proof}
As $R\leq S(0)(1-v_A(0))$, $x_H(0)$ will always be non-negative. Then, 
\begin{eqnarray*}
	\E[v_A(t+1)|v_A(t)] &=& v_A(t) \frac{v_A(t)S(t)+2c}{S(t+1)} + (1-v_A(t) ) \frac{v_A(t)S(t)}{S(t+1)} \\
		&=& v_A(t)\frac{S(t+2)}{S(t+1)}\;.
\end{eqnarray*}
It follows that 
\begin{eqnarray*}
	\E[v_A(T) ] &=& \E[v_A(T-1)]\frac{S(T+1)}{S(T)} \\
		&=& v_A(0) \frac{S(T+1)}{S(1)} 		\\
		&=& v_A(0) \Big( 1 + \frac{cT}{S(0)+c} \Big)\;.
\end{eqnarray*}



\end{document}